\documentclass{article}
\usepackage[margin=1in]{geometry}
\usepackage[english]{babel}
\usepackage[utf8]{inputenc}
\usepackage{subcaption}
\usepackage{amsmath}
\usepackage{mathtools}
\usepackage{amssymb}
\usepackage{amsfonts}
\usepackage{amsthm}
\usepackage{mathrsfs}
\usepackage[all]{xy}
\usepackage[pdftex]{graphicx}
\usepackage{color}
\usepackage{cite}
\usepackage{url}
\usepackage{indent first}
\usepackage[labelfont=bf,labelsep=period,justification=raggedright]{caption}
\usepackage[english]{babel}
\usepackage[utf8]{inputenc}
\usepackage{hyperref}
\hypersetup{
  colorlinks   = true, 
  urlcolor     = red, 
  linkcolor    = blue, 
  citecolor   = black 
}
\usepackage[colorinlistoftodos]{todonotes}
\usepackage{tkz-fct}
\usepackage{tikz}
\usepackage{authblk}
\usetikzlibrary{calc}

\title{Non-Markovian Memory-Induced Effects in Quantum Cosmology}

\begin{document}
\author[1]{Aarav Shah\thanks{\href{Email:- shahaarav103@zohomail.in}{shahaarav103@zohomail.in}}}
\author[2]{Paulo Moniz\thanks{\href{Email:-pmoniz@ubi.pt}{pmoniz@ubi.pt}}}
\author[3]{Oem Trivedi\thanks{\href{Email:- oem.trivedi@vanderbilt.edu }{oem.trivedi@vanderbilt.edu }}}
\author[4]{Meet J. Vyas\thanks{\href{Email:-a12547786@unet.univie.ac.at}{a12547786@unet.univie.ac.at}}}

\affil[1]{Virtual Institute of Astroparticle Physics, Paris 75018, France}
\affil[2]{Departamento de Física, CMA-UBI, Universidade da Beira Interior, Portugal}
\affil[3]{Department of Physics and Astronomy, Vanderbilt University, Nashville TN 37235, USA}
\affil[4]{Department of Astrophysics, University of Vienna,T\"{u}rkenschanzstra$\beta$e 17, Vienna, 1180, Austria}

\date{\today}

\maketitle

\begin{abstract}
  We study memory effects in quantum cosmology by extending the semiclassical Wheeler-DeWitt framework beyond its usual local form. The main idea is to introduce a causal memory kernel at sub leading order, rather than imposing fractional derivatives directly by hand. In this setting, fractional time evolution appears as an effective description of the underlying nonlocal dynamics. We apply the framework to cosmological perturbations in de Sitter space and find a correction to the primordial power spectrum with a characteristic $k^{3/4}$ scaling. This contribution mainly affects high $l$ CMB temperature anisotropies, in contrast with standard semiclassical quantum gravitational corrections, which are strongest at large angular scales. We also discuss how the same memory-dependent dynamics may affect primordial non-Gaussianity, producing scale dependent corrections to the bispectrum and possible deviations from the usual squeezed limit consistency relation. Since the memory coefficient controls short scale power, it may also influence structure formation and could require some tuning in order to give phenomenologically acceptable astrophysical environments. Finally, we suggest that a cyclic extension of the Hawking-Hartle no-boundary proposal may provide a setting in which the effective memory strength can evolve across successive cosmological histories. In this way, the framework gives a concrete realization of fractional quantum cosmology based on memory effects and also points to possible observational signatures of nonlocal quantum gravitational dynamics.

\end{abstract}
\section{Introduction}
The quantum description of the early universe remains one of the central challenges at the interface of gravitation, quantum theory, and cosmology \cite{bojowald2011quantum,halliwell2009introductorylecturesquantumcosmology,Wheeler:1964qna,PhysRevD.28.2960,PhysRev.160.1113,PhysRevD.30.509,Birrell:1982ix,Hawking:1973uf,Kiefer:2004xyv,Kiefer,moniz2010quantum,moniz2010quantum2,Jalalzadeh_Moniz_2022,kiefer2009cosmologicalperturbationslookclassical}. In the absence of a complete theory of quantum gravity, the Wheeler-DeWitt equation \cite{Wheeler:1964qna,PhysRev.160.1113} provides a canonical framework in which the quantum state of the universe can be studied. This has led to quantitative predictions for quantum gravitational corrections to primordial power spectra, thereby providing a potential observational window into Planck-scale physics \cite{Bojowald2021,PiazzaVareilles2025,Mazde2026,Moniz1998SUSYQC,Chataignier:2023rkq,moniz2010quantum,moniz2010quantum2,Kiefer:2011cc,PhysRevD.44.1067,Brizuela:2016gnz,Brizuela:2019jzv,Kiefer2013}. Within this setting, the semiclassical expansion developed by Kiefer and collaborators \cite{PhysRevD.93.104035,Brizuela:2016gnz,Brizuela:2019jzv,PhysRevD.44.1067,Barvinsky_1998,PhysRevD.72.045006,Kiefer:2011cc} has proven particularly powerful, it systematically recovers classical cosmological dynamics at leading order, quantum field theory on curved spacetime at next order, and controlled quantum gravitational corrections beyond.
\\
\\
In that
framework, cosmological perturbations evolve through an emergent Schr\"{o}dinger equation, with time defined by a WKB parameter constructed from the minisuperspace degrees of freedom. We summarize the main steps of this construction in Section (\ref{KK-f}). One important feature of this formalism is that it is local in the emergent time variable: both the Wheeler-DeWitt equation and the effective Schr\"{o}dinger dynamics obtained from it are local in $\eta$. There are, however, good reasons to think that this locality may only be an approximation. In quantum gravity and effective field theory, nonlocal terms often appear once some degrees of freedom are coarse-grained or integrated out \cite{biswas2013nonlocaltheoriesgravityflat,buoninfante2016ghostsingularityfreetheories,Buoninfante:2019zws,Talaganis_2015,Buoninfante_2018,Burgess_2007}. Similarly, in open quantum systems \cite{breuer2002theory}, coupling to inaccessible environmental degrees of freedom naturally produces memory effects \footnote{Throughout this manuscript, the term "memory effects" refers specifically to history dependent, temporally nonlocal effects, whereby the evolution of a system depends explicitly on its past dynamics \cite{Deser_2007,Maggiore_2014}.}, and hence non-Markovian dynamics \cite{Shrikant_2023}\footnote{Here, non-Markovianity reflects the retention of information about prior evolution rather than intrinsic information loss \cite{Shrikant_2023,PhysRevLett.103.210401,PhysRevLett.105.050403,Budini_2022}.}. If the gravitational sector effectively plays the role of such an environment for cosmological perturbations, then it is natural to expect the perturbations to retain some dependence on their past history \cite{Deser_2007,Maggiore_2014}, rather than evolving in a strictly local in-time way.
\\
\\
There is also a more specifically gravitational motivation for considering such effects. Gravitational memory is now understood to be a basic feature of gravitational physics \cite{Satishchandran_2019,strominger2014gravitationalmemorybmssupertranslations,Pate_2018,Mao_2017,Bieri_2014,Garfinkle_2017,Satishchandran_2018,Hollands_2017,Mitman_2024}: spacetime can retain a permanent imprint of earlier dynamical processes, such as the passage of gravitational radiation \cite{Zeldovich:1974gvh,PhysRevLett.67.1486,k3hl-4n82,Gasparotto:2026bru,Seto_2009,Blanchet:1992br}. These effects are not just isolated classical curiosities, since they are closely related to Bondi-Metzner-Sachs (BMS) symmetries and soft graviton theorems \cite{strominger2014gravitationalmemorybmssupertranslations}. This suggests that spacetime is not always best thought of as a purely instantaneous dynamical system, but can encode information about its previous evolution. Since such memory effects already arise in classical and semiclassical gravitational settings, it is reasonable to ask whether related, and possibly richer, memory dependence can appear in quantum gravitational regimes as well.
\\
\\
At the same time, any such memory effect should not disturb the leading semiclassical structure. The leading orders of the
expansion developed in \cite{PhysRevD.93.104035,Brizuela:2016gnz,Brizuela:2019jzv,PhysRevD.44.1067,Barvinsky_1998,PhysRevD.72.045006,Kiefer:2011cc}\footnote{For simplicity, we shall design the works by collaborators in \cite{PhysRevD.93.104035,Brizuela:2016gnz,Brizuela:2019jzv,PhysRevD.44.1067,Barvinsky_1998,PhysRevD.72.045006,Kiefer:2011cc} as the “Kiefer framework” and the order $m_P^2,m_P^0 $ and $m_P^{-2}$ equations emerging out of the semiclassical expansion as the “Kiefer equations”. See Section (\ref{KK-f}) for more details.} correctly reproduce classical background dynamics and quantum field theory on that background. This points to a simple organizing principle: memory-dependent terms, if present, should enter only as subleading $m_P^{-2}$ corrections to the local semiclassical evolution. In this view, locality is a property of the leading semiclassical limit, while memory effects represent Planck-suppressed quantum gravitational nonlocality. This perspective is also in line with recent developments in teleodynamic cosmology, where small departures from conventional thermodynamic assumptions can lead to new horizon behavior and non-equilibrium dynamics \cite{Trivedi_2026,Trivedi2026}, with memory-dependent correlations playing an important role.
\\
\\
Motivated by these points, we study memory effects in quantum cosmology and their consequences for primordial perturbations. Fractional calculus provides a natural language for such problems, because derivatives of non-integer order can encode history dependence \footnote{The application of fractional calculus to quantum cosmology has been explored in a variety of contexts in recent years \cite{Rasouli2022,Rasouli:2024crg,da_Silva_J_nior_2023,Mansouri_1999,Canedo:2025qqq,Jalalzadeh:2022uhl,Moniz:2020emn}. For a comprehensive review, see Chapter 7 of \cite{Jalalzadeh_Moniz_2022}.} \cite{Oldham1974TheFC,Samko1993FractionalIA,math13223643,liouville1832memoire,Riemann1876Versuch,podlubny1998fractional,kilbas2006theory,10.1111/j.1365-246X.1967.tb02303.x,caputo1969elasticita,Riesz1949}. However, a direct replacement of ordinary derivatives in the Wheeler-DeWitt equation, or in the Kiefer equations, by fractional derivatives turns out not to be compatible with the WKB expansion. The nonlocal nature of fractional derivatives obstructs the clean derivation of an emergent Schr\"{o}dinger equation. We therefore take a different route: memory is introduced through a nonlocal extension of the Wheeler-DeWitt equation itself, via a memory kernel appearing at subleading order. Under suitable conditions, this kernel admits an effective description in terms of fractional time derivatives, giving a controlled realization of fractional quantum cosmology rather than imposing it by hand.
\\
\\
Throughout the paper, we assume that the universe enters the classically allowed Lorentzian regime either through the Vilenkin-Yamada tunneling proposal \cite{PhysRevD.98.066003,PhysRevD.99.066010,PhysRevD.33.3560,PhysRevD.37.888,PhysRevD.30.509} or through the Hawking-Hartle no-boundary proposal \cite{PhysRevD.28.2960,Hawking1984TheQS,Hartle_Hawking_1980,Hartle_2008,hertog2023origin}. Thus the emergent conformal time begins at a finite nucleation event rather than at conformal infinity. This is important physically, since memory can accumulate only after a classical spacetime history exists, and also mathematically, since extending the lower limit of the memory integral to $\eta\rightarrow -\infty$ would generally make the resulting integrals divergent or ill-defined. In Section (\ref{frac KK-f}), we derive the corresponding modified Schr\"{o}dinger equation for perturbations and analyze it in a de Sitter background. We find that memory effects modify the scale dependence of the primordial power spectrum and leave their strongest imprint at high multipoles $l$, unlike standard semiclassical quantum gravitational corrections, which mainly affect large angular scales. This high-$l$ behavior provides a possible observational signature of memory-based quantum gravitational dynamics, and may also have consequences for small scale structure formation and the conditions needed for stable astrophysical environments.
\\
\\
In Section (\ref{Non-Gaussianity}), we extend the analysis beyond the power spectrum and investigate the possible implications of memory effects for primordial non-Gaussianity \cite{png11takahashi2014primordial,png1meerburg2019primordial,png2pajer2012new}. Since the nonlocal memory kernel can also influence higher order correlation functions, memory-induced dynamics may generate distinctive signatures in the primordial bispectrum. We show that the resulting non-Gaussian corrections inherit the blue-tilted scale dependence associated with the memory-modified power spectrum, leading to enhanced signals on small scales. In particular, departures from the standard squeezed-limit consistency relation, together with characteristic equilateral and folded bispectrum contributions, emerge as potential observational probes of non-Markovian quantum gravitational dynamics.
\\
\\
Beyond its perturbative consequences, the framework developed here also raises a broader cosmological question concerning the origin and effective magnitude of the memory strength itself. Since the memory-induced corrections predominantly modify short-scale structure formation, the corresponding memory coefficient directly influences the abundance and distribution of small scale cosmological inhomogeneities and may therefore require significant fine-tuning in order to produce the observed universe. However, within a single post-nucleation cosmological history, memory effects become physically meaningful only after the emergence of the Lorentzian regime, leaving no obvious mechanism through which the memory strength could be fine-tuned. In Section (\ref{self}), we argue that this tension may point toward a fundamentally conformal 
cyclic cosmological picture (CCC) \cite{Markwell2022TowardFA,penrose2010cycles,Penrose:2006zz,Penrose:2014vok,meissner2025physicsconformalcycliccosmology}. In particular, we explore a cyclic extension of the Hawking-Hartle no-boundary proposal in which memory-dependent correlations are inherited across successive aeons \footnote{An aeon is a complete cosmological cycle in conformal cyclic cosmology (CCC) \cite{Markwell2022TowardFA,penrose2010cycles,Penrose:2006zz,Penrose:2014vok,meissner2025physicsconformalcycliccosmology}. In CCC, the asymptotically de Sitter future of one aeon is conformally identified \cite{Penrose:1964ge} with the Big Bang hypersurface of the subsequent aeon, generating an infinite succession of cosmological aeons. }, allowing the effective memory strength to undergo cumulative inter-aeonic evolution and potentially approach phenomenologically preferred values dynamically. Among the known quantum cosmological nucleation proposals, the Hawking-Hartle framework provides the simplest setting for such a cyclic extension, since its Euclidean instanton geometry admits a comparatively straightforward reinterpretation as a conformal bridge between successive aeons. By contrast, constructing an analogous cyclic extension of the Vilenkin-Yamada tunneling proposal is considerably more subtle due to the intrinsically tunneling nature of the nucleation process. We leave a detailed investigation of cyclic tunneling cosmologies and their associated memory dynamics to future work. 
\\
\\
This work, which we discuss and summarize in Section (\ref{conc}), thus provides a bridge among semiclassical quantum cosmology, nonlocal dynamics, and fractional calculus. By grounding fractional evolution in an underlying memory kernel, this approach clarifies the origin of fractional descriptions and opens a new avenue for exploring quantum gravitational effects in the early universe.

\section{ The Kiefer 
Framework}
\label{KK-f}
Before adding memory-dependent corrections, we first recall the standard Kiefer framework, since it is the baseline for the analysis in this work. The point of reviewing it is twofold. First, it shows how the Wheeler-DeWitt equation gives rise, order by order in the Planck mass expansion, to classical background dynamics and then to a Schrodinger equation for perturbations. Second, it makes clear where the first genuine quantum gravitational corrections enter the perturbation dynamics. This ordering will be important later, because it tells us where memory-dependent terms can be added without spoiling the leading semiclassical structure.
\\
\\
We begin with the Wheeler-DeWitt equation in minisuperspace, including perturbative degrees of freedom around a homogeneous cosmological background. This equation is the starting point for the semiclassical expansion used below
\begin{equation}
    \frac{1}{2}\left\{e^{-2\alpha}\left[\frac{1}{m_P^2}\frac{\partial^2}{\partial\alpha^2}-\frac{\partial^2}{\partial \phi^2}+2e^{6\alpha}\mathcal{V}(\phi)\right]+\sum_{\mathbf{k}}\left[-\frac{\partial^2}{\partial v_{\mathbf{k}}^2}+{}^S\omega_{\mathbf{k}}^{2}(\eta)\nu_{\mathbf{k}}^2\right]+\sum_{\lambda;\mathbf{k}}\left[-\frac{\partial^2}{\partial \nu_{\mathbf{k}}^{\lambda^2}}+{}^{T}\omega_{\mathbf{k}}^2(\eta){\nu_{\mathbf{k}}^{2}}^{(\lambda)}\right]\right\}\Psi(\alpha,\phi,\{\nu_{\mathbf{k}}\})=0,
\end{equation}
where 
\begin{equation}
    \alpha=\text{ln}\left(\frac{a}{a_0}\right)
\end{equation}
and 
\begin{equation}
    m_P=\frac{3}{4\pi G},
\end{equation}
is the rescaled Planck mass ($\hbar=1$).
\\
\\
Following \cite{PhysRevD.93.104035}, we now introduce a more compact minisuperspace notation that is particularly convenient for the semiclassical expansion carried out later. In order to place the derivative terms associated with the background variables on the same footing, we first redefine the scalar field so that it becomes dimensionless,
\begin{equation}
    \tilde{\phi}=m_P^{-1}\phi.
\end{equation}
This allows us to combine the scale-factor degree of freedom and the homogeneous scalar field into a two-dimensional minisuperspace coordinate $q^A$, defined by
\begin{equation}
    q^{0}=\alpha \quad q^{1}=\tilde{\phi},
\end{equation}
where the index $A=0,1$ labels the background configuration-space variables. In terms of these variables, the minisuperspace geometry is described by the metric
\begin{equation}
    \mathcal{G}_{AB}=\text{diag}(-e^{-2\alpha},e^{2\alpha})
\end{equation}
which appears in the Wheeler-DeWitt equation through the kinetic operator $\mathcal{G}_{AB}\frac{\partial^2}{\partial q_A\partial q_B}$. We then introduce the auxiliary potential
\begin{equation}
    V(q^A)=\frac{2}{m_P^2}e^{4\alpha}\mathcal{V}(\phi).
\end{equation}
which allows the Wheeler-DeWitt equation to be written in a compact covariant form on minisuperspace.
\\
\\
Since we assume throughout this work that the cosmological perturbations remain small and that different perturbation modes evolve independently, the total wave function may be factorized into a background component and a product over individual perturbation modes,
\begin{equation}
    \Psi(\alpha,\phi,\{\nu_{\mathbf{k}}\})=\Psi_0(\alpha,\phi)\prod_{\mathbf{k};S,T_{\lambda}}\tilde\Psi_{\mathbf{k}}(\alpha,\phi,\nu_{\mathbf{k}}).
\end{equation}
This factorisation reflects the absence of mode-mode interactions at the order considered here and permits each perturbation mode to be treated separately. We therefore define, for each mode $\mathbf{k}$, an effective wave function of the form
\begin{equation}
    \Psi_{\mathbf{k}}(\alpha,\phi,\nu_{\mathbf{k}})=\Psi_0(\alpha,\nu_{\mathbf{k}})\tilde\Psi(\alpha,\phi,\nu_{\mathbf{k}}).
\end{equation}
which satisfies an individual Wheeler-DeWitt equation associated with the corresponding perturbation mode $\mathbf{k}$.
\begin{equation}
    \frac{1}{2}\left\{-\frac{1}{m_P^2}\mathcal{G}_{AB}\frac{\partial^2}{\partial q_A\partial q_B}+m_P^2V(q^A)-\frac{\partial^2}{\partial^2v_{\mathbf{k}}^2}+\omega_{\mathbf{k}}^2(\eta)\nu_{\mathbf{k}}^2\right\}\Psi_{\mathbf{k}}(\alpha,\phi,\nu_{\mathbf{k}})=0.
\end{equation}
Instead of directly solving the master Wheeler-DeWitt equation, let us apply a semi-classical approximation scheme. The advantage of
this approximation is that we recover, at consecutive orders, first the dynamics of the classical background, then
a Schr\"{o}dinger equation for the perturbations propagating on the classical background and, finally, quantum gravitational corrections to it. 
\\
\\
For our approximation, we use the following WKB-type ansatz
\begin{equation}
    \Psi_{\mathbf{k}}(q^A,v_{\mathbf{k}})=e^{iS(q^A,v_{\mathbf{k}})},
\end{equation}
where the function $S(q^A,\nu_{\mathbf{k}})$ is expanded in powers of $m_P^2$ 
in the subsequent way,
\begin{equation}
S(q^A,\nu_{\mathbf{k}})=m_P^2S_0+m_P^{0}S_1+m_P^{-2}S_2+\dots
\end{equation}
In the present case, the highest order that appears is $m_P^4$,  where we get the following equation
\begin{equation}
    \frac{\partial}{\partial v_{\mathbf{k}}}S_0(q^A,v_{\mathbf{k}})=0,
\end{equation}
which implies that the background part of the wave function represented by $S_0$ does not depend on the perturbations $v_{\mathbf{k}}$.
\\
\\
The next order is $m_P^{2}$, where we obtain the Hamilton-Jacobi equation of the background,
\begin{equation}
    \mathcal{G}_{AB}\frac{\partial S_0}{\partial q_A}\frac{\partial S_0}{\partial q_B}+V(q^A)=0.
\end{equation}
It can be shown that this equation is equivalent to the Friedmann equation.
\\
\\
The subsequent order, which is $m_P^0$, yields the equation
\begin{equation}\label{m_P^0}
    2\mathcal{G}_{AB}\frac{\partial S_0}{\partial q_A}\frac{\partial S_1}{\partial q_B}-i\mathcal{G}_{AB}\frac{\partial^2 S_0}{\partial q_A\partial q_B}+\left(\frac{\partial S_1}{\partial \nu_{\mathbf{k}}}\right)^2-i\frac{\partial^2S_1}{\partial \nu_{\mathbf{k}}^2}+\omega_{\mathbf{k}}^2\nu_{\mathbf{k}}^2=0.
\end{equation}
As we can see, the perturbations represented by $S_1$ and $\mathcal{\nu}_{\mathbf{k}}$  now come into play. In order to obtain a Schr\"{o}dinger equation  for the perturbations modes, we define a wave function $\psi_{\mathbf{k}}^{(0)}$ in the following way
\begin{equation}
    \psi_{\mathbf{k}}^{(0)}(q^A,\nu_{\mathbf{k}})=\gamma(q^A)e^{iS_1(q^A,\nu_{\mathbf{k}})}.
\end{equation}
Here we have introduced a prefactor $\gamma$, which is the inverse of the standard WKB prefactor, which we did not include in our WKB-type ansatz. We demand that $\gamma$ obey the following condition
\begin{equation}
    \mathcal{G}_{AB}\frac{\partial}{\partial q_A}\left[\frac{1}{2\gamma^2}\frac{\partial S_0}{\partial q_B}\right]=0.
\end{equation}
Additionally, we can define the conformal WKB time, which we will identify with the classical conformal time, in terms of the minisuperspace variables as follows
\begin{equation}
    \frac{\partial }{\partial \eta}=\mathcal{G}_{AB}\frac{\partial S_0}{\partial q_A}\frac{\partial }{\partial q_B}=e^{-2\alpha}\left[-\frac{\partial S_0}{\partial \alpha}\frac{\partial }{\partial \alpha}+\frac{\partial S_0}{\partial \tilde\phi}\frac{\partial}{\partial\tilde\phi}\right].
\end{equation}
Using these relations, it is possible to rewrite eqn.~\eqref{m_P^0} for $S_1$ as a Schr\"{o}dinger equation for $\psi^{(0)}_{\mathbf{k}}$,
\begin{equation}\label{mp0 schrodinger}
\mathcal{H}_{\mathbf{k}}\psi^{(0)}_{\mathbf{k}}=i\frac{\partial}{\partial \eta}\psi_{\mathbf{k}}^{(0)},
\end{equation}
where the perturbation Hamiltonian is given by 
\begin{equation}
    \mathcal{H}_{\mathbf{k}}=-\frac{1}{2}\frac{\partial^2}{\partial\nu_{\mathbf{k}}^2}+\frac{1}{2}\omega_{\mathbf{k}}^2\nu_{\mathbf{k}}^2.
\end{equation}
Therefore, we end up with a Schr\"{o}dinger equation for the quantum states of the perturbations, where the time is defined from the minisuperspace quantum dynamics.
\\
\\
Up to now, we have recovered well-known physics,
so the interesting part and main point of this investigation comes at the next order $m_P^{-2}$, where quantum gravitational effects come into play. At this order, the equation we have to consider is given by
\begin{equation}
    \mathcal{G}_{AB}\frac{\partial S_0}{\partial q_A}\frac{\partial S_2}{\partial q_B}+\frac{1}{2}\mathcal{G}_{AB}\frac{\partial S_1}{\partial q_A}\frac{\partial S_1}{\partial q_B}-\frac{i}{2}\mathcal{G}_{AB}\frac{\partial^2 S_1}{\partial q_A\partial q_B}+\frac{\partial S_1}{\partial \nu_{\mathbf{k}}}\frac{\partial S_2}{\partial \nu_{\mathbf{k}}}-\frac{i}{2}\frac{\partial^2S_2}{\partial \nu_{\mathbf{k}}^2}=0
\end{equation}
The newly appearing function $S_2$,  which contains information about the quantum gravitational corrections at this order, is then split into a part $\zeta$ that depends only
on the minisuperspace variables, and a part $\chi$ that contains also the perturbations $\nu_{\mathbf{k}}$,
\begin{equation}
    S_2(q^A,\nu_{\mathbf{k}})=\zeta(q^A)+\chi(q^A,\nu_{\mathbf{k}}).
\end{equation}
The reason for this split is to isolate the next-order correction to the WKB prefactor, which is given by $\zeta$ and just contributes a phase. After this split, we end up with the subsequent equation for $\chi$,
\begin{equation}
    \frac{\partial \chi}{\partial \eta}=\frac{1}{\psi_{\mathbf{k}}^{(0)}}\left(-\frac{1}{\gamma}\mathcal{G}_{AB}\frac{\partial \psi_{\mathbf{k}}^{(0)}}{\partial q_A}\frac{\partial \gamma}{\partial q_B}+\frac{1}{2}\mathcal{G}_{AB}\frac{\partial^2\psi_{\mathbf{k}}^{(0)}}{\partial q_A\partial q_B}+i\frac{\partial \psi_{\mathbf{k}}^{(0)}}{\partial \nu_{\mathbf{k}}}\frac{\partial \chi}{\partial \nu_{\mathbf{k}}}+\frac{i\psi_{\mathbf{k}}}{2}\frac{\partial^2\chi}{\partial \nu_{\mathbf{k}}^2}\right).
\end{equation}
We can now use $\chi$ to define a new function $\psi_{\mathbf{k}}^{(1)}$ that contains the quantum gravitational corrections of
the order $m_P^{_2}$ as follows 
\begin{equation}
    \psi_{\mathbf{k}}^{(1)}(q^A,\nu_{\mathbf{k}})=\psi_{\mathbf{k}}^{(0)}(q^A,\nu_{\mathbf{k}})e^{im_P^{-2}\chi(q^A,\nu_{\mathbf{k}})}.
\end{equation}
Using this definition, we obtain a Schr\"{o}dinger equation for $\psi^{(1)}_{\mathbf{k}}$ with a quantum gravitational correction term
that is suppressed by a prefactor of $m_p^{-2}$
\begin{equation}
    i\frac{\partial}{\partial \eta}\psi_{\mathbf{k}}^{(1)}=\mathcal{H}_{\mathbf{k}}\psi^{(1)}_{\mathbf{k}}+\frac{\psi_{\mathbf{k}}^{(1)}}{m_P^2\psi^{(0)}_{\mathbf{k}}}\left(\frac{1}{\gamma}\mathcal{G}_{AB}\frac{\partial \psi_{\mathbf{k}}^{(0)}}{\partial q_A}\frac{\partial \gamma}{\partial q_B}-\frac{1}{2}\mathcal{G}_{AB}\frac{\partial^2\psi_{\mathbf{k}}^{(0)}}{\partial q_A\partial q_B}\right).
\end{equation}
In terms of the perturbation Hamiltonian, we obtain the following form of the QG-corrected Schr\"{o}dinger equation
\begin{equation}\label{QG CORRECTED SCHRODING PERURBATION}
  i\frac{\partial}{\partial \eta}\psi_{\mathbf{k}}^{(1)}=\mathcal{H}_{\mathbf{k}}\psi^{(1)}_{\mathbf{k}}-\frac{\psi_{\mathbf{k}}^{(1)}}{m_P^2\psi^{(0)}_{\mathbf{k}}}\left[ \frac{(\mathcal{H}_\mathbf{k})^2}{V}\psi_{\mathbf{k}}^{(0)}+i\frac{\partial}{\partial \eta}\left(\frac{\mathcal{H}_{\mathbf{k}}}{V}\right) \psi_{\mathbf{k}}^{(0)}\right]
\end{equation}
We shall now make a Gaussian ansatz to find the solution for the $m_P^{0}$ order Schr\"{o}dinger equation \eqref{mp0 schrodinger}. Introducing a normalization factor $N^{(0)}_{\mathbf{k}}$ and a Gaussian width $\Omega^{(0)}_{\mathbf{k}}(\eta)$, our ansatz reads
\begin{equation}
    \psi_{\mathbf{k}}^{(0)}(\eta,\nu_{\mathbf{k}})=N_{\mathbf{k}}^{(0)}(\eta)e^{-\frac{1}{2}\Omega^{(0)}_{\mathbf{k}}(\eta)\nu_{\mathbf{k}}^2}.
\end{equation}
Inserting this ansataz to eqn.~\eqref{mp0 schrodinger} and collecting all the terms with either a factor of $\nu_{\mathbf{k}}^2$ or $\nu_{\mathbf{k}}^{0}$, we arrive at 
\begin{equation}
    iN^{(0)}_{\mathbf{k}}(\eta)'=\frac{1}{2}N_{\mathbf{k}}^{(0)}(\eta)\Omega_{\mathbf{k}}^{(0)}(\eta),
\end{equation}
\begin{equation}\label{Gaussian width}
    i\Omega_{\mathbf{k}}^{(0)}(\eta)'=(\Omega_{\mathbf{k}}^{(0)}(\eta))^2-\omega_{\mathbf{k}}(\eta).
\end{equation}
Furthermore, we require a normalized wave function for any given time $\eta$
\begin{equation}
    |\psi_{\mathbf{k}}^{(0)}|^2=|N_{\mathbf{k}}^{(0)}|^2\frac{\sqrt{\pi}}{\sqrt{\mathfrak{Re}\Omega_{\mathbf{k}}^{(0)}}}=1.
\end{equation}
This provides the modulus of the normalization factor in terms of the real part of the Gaussian width. Therefore, we arrive at the following relation
\begin{equation}\label{N_k^0 1/4}
    N^{(0)}_{\mathbf{k}}=\left(\frac{\mathfrak{Re}\Omega_{\mathbf{k}}^{(0)}}{\pi}\right)^{1/4}e^{i\phi},
\end{equation}
where $\phi$ is a real function of time $\eta$. Eqn.~\eqref{Gaussian width} is a Riccati equation and, as is well known, can always be written as a second order differential equation. For this purpose, the following change of variable is performed,
\begin{equation}
    \Omega_{\mathbf{k}}^{(0)}(\eta)=-i\frac{y_{\mathbf{k}}^{(0)}(\eta)'}{y_{\mathbf{k}}^{(0)}(\eta)},
\end{equation}
which leads to the equation of a  parametric oscillator,
\begin{equation}
    y_{\mathbf{k}}^{(0)}(\eta)''+\omega_{\mathbf{k}}^2(\eta)y_{\mathbf{k}}^{(0)}(\eta)=0.
\end{equation}
Let us now study the same Gaussian ansatz  for the quantum gravitationally corrected Schr\"{o}dinger equation
\begin{equation}
      \psi^{(1)}(\eta,\nu_{\mathbf{k}})=N_{\mathbf{k}}^{(1)}(\eta)e^{-\frac{1}{2}\Omega_{\mathbf{k}}^{(1)}\nu_{\mathbf{k}}^2}.
\end{equation}
Inserting this ansatz into eqn.~\eqref{QG CORRECTED SCHRODING PERURBATION} and collecting all the terms with either a factor of $\nu_{\mathbf{k}}^2$ or $\nu_{\mathbf{k}}^{0}$, we arrive at 
\begin{equation}
        iN_{\mathbf{k}}^{(1) '}(\eta)=\frac{1}{2}N_{\mathbf{k}}^{(1)}(\eta)\Omega_{\mathbf{k}}^{(1)}(\eta), 
\end{equation}
\begin{equation}
    i\Omega_{\mathbf{k}}^{(1)'}(\eta)=\left(\Omega_{\mathbf{k}}^{(1)}(\eta)\right)^2-\tilde{\omega_{\mathbf{k}}}^2(\eta),
\end{equation}
where $\tilde{\omega_{\mathbf{k}}}$ is given by
\begin{equation}
    \tilde{\omega_{\mathbf{k}}}^2=\omega_{\mathbf{k}}-\frac{1}{2m_P^2V}\left[\left(3\Omega_{\mathbf{k}}^{(0)}-i(\text{ln}V)\right)'\left(\omega_{\mathbf{k}}^2-\left(\Omega_{\mathbf{k}}^{(0)}(\eta)\right)^2\right)+2i\omega_{\mathbf{k}}\omega_{\mathbf{k}}'\right].
\end{equation}
The differential structure of the equation is the same, and thus we can perform the same change of variable
\begin{equation}
    \Omega_{\mathbf{k}}^{(0)}(\eta)=-i\frac{y_{\mathbf{k}}^{(1)}(\eta)'}{y_{\mathbf{k}}^{(1)}(\eta)},
\end{equation}
to obtain a linear equation for the auxiliary variable $y_{\mathbf{k}}^{(1)}$,
\begin{equation}
    y_{\mathbf{k}}(\eta)''+\tilde{\omega}^2_{\mathbf{k}}(\eta)y_{\mathbf{k}}(\eta)=0.
\end{equation}
Let us now focus on a de Sitter background and try to obtain solutions for the uncorrected and corrected Schr\"{o}dinger equation in that case.
\\
\\
We construct a de Sitter universe by setting the scalar field $\phi$ to a constant value. We can write the following relation between the potential $\mathcal{V}(\phi)$ of the scalar field and the constant Hubble parameter $H_0$ in our de Sitter universe,
\begin{equation}
    \mathcal{V}(\phi)=\frac{3}{8\pi G}H_0^2=\frac{1}{2}m_P^2H_0^2.
\end{equation}
In this background, the effective frequencies $\omega_{\mathbf{k}}^2(\eta)$  for scalar and tensor perturbations take the same
form for both scalar and tensor perturbations,
\begin{equation}
    \omega_{\mathbf{k}}^2(\eta)=k^2-\frac{2}{\eta^2}.
\end{equation}
so that both sectors obey the same mode equation.
\\
\\
The Hamilton-Jacobi equation for the background geometry then reduces to
\begin{equation}
    \left(\frac{\partial S_0}{\partial \alpha}\right)^2-e^{6\alpha}H_0^2=0.
\end{equation}
whose solution is
\begin{equation}
    S_0=-\frac{1}{3}e^{3\alpha}H_0.
\end{equation}
This solution determines the semiclassical background trajectory and fixes the corresponding WKB time evolution.
\\
\\
Substituting the de Sitter frequency into the Riccati equation \eqref{Gaussian width} for the Gaussian width yields
\begin{equation}
    i\Omega_{\mathbf{k}}^{(0)}(\eta)'=\left(\Omega_{\mathbf{k}}^{(0)}(\eta)\right)^2-k^2+\frac{2}{\eta^2}.
\end{equation}
As discussed previously, this nonlinear equation can be transformed into a linear second order differential equation through the auxiliary variable $y_{\mathbf{k}}^{(0)}$. The resulting mode function takes the general form
\begin{equation}
    y_{\mathbf{k}}^{(0)}(\eta)=A(k)e^{-ik\eta}\left(1-\frac{i}{k\eta}\right)+B(k)e^{ik\eta}\left(1+\frac{i}{k\eta}\right).
\end{equation}
where $A(k)$ and $B(k)$ are integration constants determined by the choice of initial quantum state.
\\
\\
To recover the standard adiabatic vacuum at short wavelengths, we impose the Bunch-Davies boundary condition. Physically, this corresponds to requiring that sufficiently sub-horizon modes behave like ordinary Minkowski vacuum fluctuations at early times. This selects the positive-frequency solution
\begin{equation}
     y_{\mathbf{k}}^{(0)}(\eta)=\frac{1}{\sqrt{2k}}e^{ik\eta}\left(1+\frac{i}{k\eta}\right).
\end{equation}
from which the Gaussian width becomes
\begin{equation} \label{Omega 0 desitter}
    \Omega_{\mathbf{k}}^{(0)}(\eta)=\frac{k^3\eta^2}{1+k^2\eta^2}+\frac{i}{\eta(1+k^2\eta^2)}.
\end{equation}
Having obtained the leading-order solution, we now turn to the quantum gravitationally corrected Schr\"{o}dinger equation. Since the corrections are suppressed by $m_P^{-2}$,
it is natural to treat them perturbatively by decomposing the Gaussian width into an uncorrected part and a small correction,
\begin{equation}
\Omega_{\mathbf{k}}^{(1)}=\Omega_{\mathbf{k}}^{(0)}+\tilde\Omega_{\mathbf{k}}^{(1)}.
\end{equation}
If the function $\Omega_{\mathbf{k}}^{(1)}$ is assumed to be small, 
we can drop its quadratic terms. In this way, we have
\begin{equation}\label{diff eqaution qg corrected omega}
    i\tilde\Omega_{\mathbf{k}}^{(1)}=2\Omega_{\mathbf{k}}^{(0)}\Omega_{\mathbf{k}}^{(1)}-(\tilde\omega_{\mathbf{k}}^2-\omega_{\mathbf{k}}^2).
\end{equation}
Explicitly, eqn.~\eqref{diff eqaution qg corrected omega} reads
\begin{equation}
    i\tilde\Omega_{\mathbf{k}}^{(1) \space'}=\frac{2k^3\eta^3+2i}{\eta(1+k^2\eta^2)}\tilde\Omega_{\mathbf{k}}^{(1)}+\frac{H_0^2\eta^4}{2m_P^2}\frac{k^3(11-k^2\eta^2)}{(1+k^2\eta^2)^3}.
\end{equation}
Its general solution is given by
\begin{equation}
    \tilde\Omega_{\mathbf{k}}^{(1)}=-\frac{\eta^2e^{-2i\eta k}}{(\eta k+i)^2}\left\{c_1+\frac{H_0^2}{4m_P^2}\left[9e^{-2}\Gamma(0,-2ik\eta-2)+3e^2\Gamma(0,2-2ik\eta)-e^{2i\eta k}\frac{1+\eta k(\eta k+6i)}{(\eta k-i)^2}\right]\right\},
\end{equation}
where $c_1$ is an integration constant fixed by the choice of initial conditions.
\\
\\
The quantum gravitationally corrected power spectra are then given by
\begin{equation}
  \mathcal{P}_{S,T}^{(1)}(k)=\mathcal{P}_{S,T}^{(0)}(k)\frac{\mathfrak{Re}\Omega_{\mathbf{k}}^{(1)}}{\mathfrak{Re}\Omega_{\mathbf{k}}^{(0)}}=\mathcal{P}_{S,T}^{(0)}(k)\left[1+0.988\frac{H_0}{m_P^2}\left(\frac{k_0}{k}\right)^3\right],
\end{equation}
where $k_0$ is a reference wave
number that represents the inverse of the length scale. Note that standard Quantum Gravitational effects induce an additional $k^{-3}$ dependence, which renders the corrected power spectra become explicitly
scale-dependent and enhanced at largest scales. Since primordial perturbations directly source the temperature anisotropies of the cosmic microwave background \cite{Sachs:1967er,1992ApJ...396L...1S,Kogut_1993,Hu_1995,Hu_2002,2014PTEP.2014fB101S,2020v,2020vii}, this scale dependence propagates into the CMB temperature anisotropy $C_l$, 
\begin{equation}
    C_l^{(i)}=\int_0^{\infty} \frac{dk}{k}\mathcal{P}_{S}^{(i)}(k)\Theta_l^2(k),
\end{equation}
where $i=0,1$ denotes the uncorrected and corrected coefficients, respectively, and $\Theta_l(k)$ is the transfer function. Now, for large scales (small $l$), the transfer function can be written in terms of the spherical Bessel functions $j_l$ as
\begin{equation}
    \Theta_l(k)=\frac{1}{3}j_l(k[\eta_{\text{hor}}-\eta_{\text{rec}}]),
\end{equation}
where $\eta_{\text{rec}}$ is the conformal time at horizon crossing and $\eta_{\text{rec}}$ is the conformal time at recombination.
\\
\\
Let us define the quantum gravitational correction to
the temperature anisotropies in the following way,
\begin{equation}
    \Delta C_l=C_l^{(1)}-C_l^{(0)}.
\end{equation}
Applying the results for the corrected scalar power spectrum, we get that, for large scales, this correction has the
following form,
\begin{equation}
    \Delta C_l\simeq \frac{1}{4\pi^2}\int_0^{\infty}\frac{dk}{k \epsilon}\left(\frac{H_k}{m_P}\right)^4\left(\frac{\bar{k}}{k}\right)^3j_l^2(k[\eta_{\text{hor}}-\eta_{\text{rec}}].
\end{equation}
Now, the Bessel function is strongly peaked around $k|\eta_{\text{hor}}-\eta_{\text{rec}}|\simeq l$ and effectively acts as a Dirac delta function mapping $k$ to $l$. Therefore, we can integrate the explicit $k$ dependence, leading to 
\begin{equation}
    \Delta C_l\simeq \frac{3}{4\pi \epsilon}\left(\frac{H_k}{m_P}\right)^4\frac{|\bar{k}(\eta_{\text{hor}}-\eta_{\text{rec}})|^3}{(2l-3)(2l-1)(2l+1)(2l+3)(2l+5)}.
\end{equation}
Hence, standard quantum gravitational effects produce their most significant modifications in the low $l$ sector of the CMB anisotropy spectrum (corresponding to large angular scales on the sky) and are strongly suppressed in the high $l$ sector (corresponding to small angular scales on the sky).

\section{Fractionating the Kiefer
Framework}
\label{frac KK-f}
The semiclassical expansion of the Wheeler-DeWitt equation, as reviewed in Section (\ref{KK-f}),
gives quantum gravitational corrections to the Schrodinger dynamics of cosmological perturbations at order $m_P^{-2}$. In the standard treatment these corrections are local in the emergent WKB time $\eta$, as expected from the local structure of the Wheeler-DeWitt equation in minisuperspace. From a broader quantum gravitational point of view, however, there is no reason why the effective dynamics of perturbations must remain exactly local in time once inaccessible gravitational degrees of freedom, environmental variables, or high energy modes have been coarse-grained away. Such procedures naturally lead to memory terms and nonlocal evolution equations, as is familiar from non-Markovian open quantum systems \cite{Shrikant_2023,PhysRevLett.103.210401,PhysRevLett.105.050403,Budini_2022} and from effective nonlocal descriptions of quantum gravity \cite{biswas2013nonlocaltheoriesgravityflat,buoninfante2016ghostsingularityfreetheories,Buoninfante:2019zws,Talaganis_2015,Buoninfante_2018,Burgess_2007}. This motivates us to ask how memory-dependent corrections can be introduced into the Kiefer
framework without destroying its semiclassical structure, and whether such corrections can, under suitable assumptions, be described effectively by fractional time derivatives.
\\
\\
The most direct idea would be to insert fractional derivatives into the Wheeler-DeWitt equation itself. For example, one could try to replace the local differential operators in the minisuperspace formulation by Caputo fractional derivatives, since these are often used to describe nonlocal time evolution with memory. This route is attractive at first sight because fractional derivatives provide a compact way of representing history dependence, with the fractional order measuring the strength of the memory effect. However, as we will see, this direct procedure is too naive for the semiclassical Wheeler-DeWitt expansion. The WKB construction depends crucially on the ordinary derivative structure of the equation, and replacing those derivatives by fractional ones obstructs the clean separation of orders needed to recover the background Hamilton-Jacobi equation and the perturbative Schr\"{o}dinger equation.
\\
\\
Following this idea, we may formally replace the second order derivatives in the Wheeler-DeWitt equation by fractional derivatives of order $\alpha$, leading to 
\begin{equation}
        \frac{1}{2}\left\{-\frac{1}{m_P^2}\mathcal{G}_{AB}{}^{C}D^{\alpha}_{q_A}{}^{C}D^{\alpha}_{q_B}+m_P^2V(q^A)- {}^{C}D^{2\alpha}_{v_{\mathbf{k}}}+\omega_{\mathbf{k}}^2(\eta)\nu_{\mathbf{k}}^2\right\}\Psi_{\mathbf{k}}(\alpha,\phi,\nu_{\mathbf{k}})=0.
\end{equation}
Here, ${}^{C}D^{\alpha}$denotes the Caputo derivative \cite{10.1111/j.1365-246X.1967.tb02303.x,caputo1969elasticita} \footnote{We use Caputo \cite{10.1111/j.1365-246X.1967.tb02303.x,caputo1969elasticita} rather than Riemann-Lioville fractional derivatives \cite{Riemann1876Versuch,liouville1832memoire} because the Caputo formalism allows initial conditions to be specified in terms of ordinary integer-order derivatives of the wave function.} at fractional order $\alpha$.
However, this naive fractionalization encounters serious difficulties when we attempt to implement the semiclassical expansion underlying the Kiefer framework.
The standard derivation relies on the WKB-type ansatz
\begin{equation}
    \Psi_{k}=\text{exp}(iS(q^A,\nu_{\mathbf{k}})),
\end{equation}
together with an expansion of $S$ in the powers of $m_{P}^{-2}$. Crucially, this procedure depends on the fact that ordinary derivatives act on the exponential in a simple multiplicative way, allowing a consistent separation of orders and leading to the Hamilton-Jacobi equation for the background and a Schr\"{o}dinger equation for perturbations.
\\
\\
In contrast, fractional derivatives, such as the Caputo derivative, are nonlocal operators and do not satisfy the standard Leibniz or chain rules. As a result, their action on the exponential ansatz does not yield a closed expression in terms of derivatives of $S_0,S_1,S_2,\dots$. Instead, we obtain nonlocal integral expressions involving the full history of the wave function, which prevents a systematic expansion in powers of $m_P^{-2}$.
\\
\\
Alternatively, we may fractionate the Kiefer equations as follows,
\begin{flalign}
    & \mathcal{G}_{AB}{}^CD^{\alpha}_{q_A}S_0{}^CD^{\alpha}_{q_B}S_0+V(q^A)=0,\\
    & 2\mathcal{G}_{AB}{}^CD^{\alpha}_{q_A}S_0{}^CD^{\alpha}_{q_B}S_1-i{}^{C}D^{\alpha}_{q_A}{}^{C}D^{\alpha}_{q_B}S_0+\left({}^{C}D_{\mathcal{\nu}_{\mathbf{k}}}S_1\right)^2-i{}^{C}D^{2\alpha}_{\nu_{\mathbf{k}}}S_1+\omega_{\mathbf{k}}^2\nu_{\mathbf{k}}^2=0,\\
    & \mathcal{G}_{AB}{}^{C}D^{\alpha}_{q_A}S_0{}^{C}D^{\alpha}_{q_B}S_2+\frac{1}{2}\mathcal{G}_{AB}{}^{C}D^{\alpha}_{q_A}S_1{}^{C}D^{\alpha}_{q_B}S_1-\frac{i}{2}\mathcal{G}_{AB}{}^{C}D^{\alpha}_{q_A}{}^{C}D^{\alpha}_{q_B}S_1+{}^{C}D^{\alpha}_{\nu_{\mathbf{k}}}S_1{}^{C}D^{\alpha}_{\nu_\mathbf{k}}S_2-\frac{i}{2}{}^{C}D^{2\alpha}_{\nu_{\mathbf{k}}}S_2=0,
\end{flalign}
and define the conformal time as
\begin{equation}
    ^{C}D^{\alpha}_{\eta}=\mathcal{G}_{AB}{}^{C}D^{\alpha}_{q_A}S_0{}^{C}D^{\alpha}_{q_B}.
\end{equation} 
Even now, we cannot obtain a closed Sch\"{o}dinger type equation. At best, we can formally rewrite the resulting equation in terms of logarithmic derivatives of the wave function, leading to expressions that involve nested fractional derivatives acting on both background and perturbation variables. Concretely,
\begin{equation}\label{Messy and naive schrodinger}
\begin{split}
    i\space {}^{C}D^{\alpha}_{\eta} (\text{ln}\psi^{1})=\frac{({}^{C}D^{\alpha}_{\nu_{\mathbf{k}}}S_1)^2-i{}^{C}\space D^{2\alpha}_{\nu_{\mathbf{k}}}S_1+\omega_{\mathbf{k}}^2\nu_{\mathbf{k}}^2}{2}\\+\frac{1}{m_P^2}\left[
    \begin{split}
    \frac{-1}{2}\mathcal{G}_{AB}{}^C{D}^{\alpha}_{q_A}\text{ln}\left(\frac{\psi_{\mathbf{k}}^{0}}{\gamma(q^A)}\right){}^CD^{\alpha}_{q_B}\text{ln}\left(\frac{\psi_{\mathbf{k}}^{0}}{\gamma(q^B)}\right)-\frac{1}{4}\mathcal{G}_{AB}{}^{C}D^{\alpha}_{q_A}{}^{C}D^{\alpha}_{q_B}\text{ln}\left(\frac{\psi_{\mathbf{k}}^{0}}{\gamma(q^B,q^A)}\right)\\-\frac{i}{2}{}^{C}D^{\alpha}_{\nu_\mathbf{k}}(\text{ln}(\psi_{\mathbf{k}}^{0})){}^{C}D^{\alpha}_{\nu_{\mathbf{k}}}\chi -\frac{i}{2}{}^{C}D^{2\alpha}_{\nu_{\mathbf{k}}}\chi
    \end{split}
    \right].
\end{split}
\end{equation}
Eqn.~\eqref{Messy and naive schrodinger} is highly nonlocal and does not admit a clear probabilistic interpretation or even a closed Schr\"{o}dinger-type form.
This indicates that a direct replacement of derivatives by fractional derivatives is not compatible with the semiclassical structure of the Kiefer
expansion.
\\
\\
The failure of the naive fractionalization procedure indicates that nonlocality should not be introduced in an \textit{ad hoc} manner at the level of differential operators. Instead, we take the view that memory effects are a fundamental feature of quantum gravitational dynamics and should therefore be incorporated directly into the Wheeler-DeWitt equation itself.
\\
\\
From this perspective, the standard local Wheeler-DeWitt equation should be regarded as an approximation, valid in regimes where such memory effects can be neglected. More generally, we expect the quantum state of the universe to exhibit nonlocal correlations in the emergent time, leading to an intrinsically non-Markovian structure. We therefore introduce a memory-dependent modification of the Wheeler-DeWitt equation, in which the evolution at a given time $\eta$ depends explicitly on its past history through a nonlocal kernel.
\\
\\
Since the standard Kiefer
expansion shows that quantum gravitational corrections first appear at order $m_P^{-2}$, it is reasonable 
to introduce memory effects at the same order. In this way, nonlocality is treated as a subleading correction to the local Wheeler-DeWitt dynamics, ensuring consistency with the semiclassical hierarchy. Now a generic memory correction of a function $h(x)$ induces an additional term to its derivative,
\begin{equation}
\frac{dh}{dx}\bigg|_{\text{modified}} = \frac{dh}{dx}\bigg|_{\text{standard}} + \int_{a}^{x}K(x-x')h(x') \,dx',
\end{equation}
where $K(x-x')$ is known as the memory kernel and is a distribution encoding the nonlocal influence of values of $x'<x$ on the function $h(x)$. 
\\
\\
Similarly, we expect that a general memory correction to the Wheeler-DeWitt equation can be written as an integral over past times,
\begin{equation}
   \frac{1}{m_P^2} \int_{0}^{\eta} K(\eta-\eta') \frac{\partial \Psi(\eta')}{\partial \eta'} d\eta'.
\end{equation}
where $K(\eta-\eta')$ is the memory kernel encoding the influence of past configurations. Here we work within the Vilenkin-Yamada tunneling proposal, although the same reasoning applies equally well to the Hawking-Hartle no-boundary proposal; for convenience, we set the nucleation time to $\eta_{\text{nuc}}=0$. The memory kernel is assumed to encode the cumulative influence of prior evolution, analogous to non-Markovian memory in open quantum systems. Such a notion becomes physically meaningful only after the universe has emerged into the classically allowed Lorentzian regime, where the WKB parameter $\eta$ is interpreted as physical conformal time and the universe follows a well-defined causal trajectory. In the classically forbidden region preceding nucleation, the universe is described by an under-the-barrier wave function rather than by a classical spacetime history, so there is no meaningful temporal evolution along which information can accumulate. Consequently, the memory integral begins at the nucleation event and includes only the post-nucleation evolution of the universe. This choice is also mathematically advantageous. Although one often formally treats the classical solution as if it originated from $\eta\rightarrow -\infty$, adopting $-\infty$ as the lower limit of the memory integral would generally render the subsequent integrals divergent or ill-defined, since the kernel would accumulate contributions from an infinitely extended past. Starting the integration at $\eta_{\text{nuc}}=0$ therefore has both a clear physical interpretation and ensures that the resulting memory corrections remain finite and well defined.
\\
\\
The modified Wheeler-DeWitt equation then takes the form
\begin{equation}\label{Memory Wheeler De Witt}
    \frac{1}{2}\left\{-\frac{1}{m_P^2}\mathcal{G}_{AB}\frac{\partial^2}{\partial q_A\partial q_B}+m_P^2V(q^A)-\frac{\partial^2}{\partial^2v_{\mathbf{k}}^2}+\omega_{\mathbf{k}}^2(\eta)\nu_{\mathbf{k}}^2\right\}\Psi_{\mathbf{k}}(\alpha,\phi,\nu_{\mathbf{k}})=\frac{1}{m_P^2} \int_0^{\eta} K(\eta-\eta') \frac{\partial \Psi(\eta')}{\partial \eta'} d\eta'.
\end{equation}
The right-hand side introduces a nonlocal correction to the Wheeler-DeWitt equation, such that the evolution at time $\eta$ depends on the entire past history of the wave function. This structure is characteristic of non-Markovian dynamics and is analogous to memory kernels that appear in effective descriptions of open quantum systems. Since the memory term is suppressed by $m_P^{-2}$, the leading orders of the semiclassical expansion remain unchanged. In particular, the Hamilton-Jacobi equation for the background and the Schr\"{o}dinger equation for $\psi^{(0)}_{\mathbf{k}}$ are recovered as in the standard Kiefer
framework.
\\
\\
The QG Schr\"{o}dinger equation for $\psi^{(1)}_{\mathbf{k}}$ is now corrected as,
\begin{equation} \label{effective memory schrodinger}
        i\int\delta(\eta-\eta')\frac{\partial}{\partial \eta'}\psi_{\mathbf{k}}^{(1)}(\eta')d\eta'=\mathcal{H}_{\mathbf{k}}\psi^{(1)}_{\mathbf{k}}+\frac{\psi_{\mathbf{k}}^{(1)}}{m_P^2\psi^{(0)}_{\mathbf{k}}}\left(\frac{1}{\gamma}\mathcal{G}_{AB}\frac{\partial \psi_{\mathbf{k}}^{(0)}}{\partial q_A}\frac{\partial \gamma}{\partial q_B}-\frac{1}{2}\mathcal{G}_{AB}\frac{\partial^2\psi_{\mathbf{k}}^{(0)}}{\partial q_A\partial q_B}+\int_0^{\eta} K(\eta-\eta') \frac{\partial \Psi}{\partial \eta'} d\eta'\right).
\end{equation}
We now specify the form of the memory kernel that enters the nonlocal correction. Since the memory term is introduced only at order $m_P^{-2}$, the kernel has to satisfy a few basic conditions. First, it must be causal, so that the evolution at time $\eta$ depends only on the earlier interval $0\leq \eta' < \eta$. Second, it should have the right short-distance behavior near $\eta'=\eta$, so that the local first-order time derivative is recovered when memory effects are removed, while the integral still acts as a well defined distribution on smooth functions. Third, at leading order in the memory strength, it should produce an effective fractional time evolution whose order differs from unity only by a Planck suppressed amount. These conditions naturally lead us to use a scale-free plus-distribution kernel. Such a kernel can describe weak long-time memory without adding a new external time scale.
\\
\\
There is also a physical reason to allow the memory strength to depend on the perturbation mode. If the kernel is meant to encode the accumulated influence of the cosmological history on perturbations, then different Fourier modes need not remember that history in exactly the same way. Modes with different wavelengths cross the horizon at different times and sample different parts of the background evolution. It is therefore reasonable for the memory kernel to retain some information about scale, instead of being completely universal in $k$. This motivates the inclusion of a scale-tracking prefactor proportional to the ratio of mode amplitudes $\psi_{\mathbf{k}}^{(1)}/\psi_{\mathbf{k_0}}^{(1)}$. We shall therefore take the memory distribution to have the form
\begin{equation} \label{kernel}
K(\eta-\eta')= \frac{-iA\psi_{\mathbf{k}}^{(1)}(\eta)}{\psi_{\mathbf{k_0}}^{(1)}(\eta)} \left[\frac{1}{\eta-\eta'}\right]_{+},
\end{equation}
where $A$ parametrizes the strength of the memory effect, the symbol $[\cdot]_{+}$ denotes the Hadamard finite part/ plus distribution. Explicitly for a sufficiently regular function $f(\eta')$, it is defined through
\begin{equation}
\int_{0}^{\eta} \left[\frac{1}{\eta-\eta'}\right]_{+} f(\eta') \, d\eta' = \int_0^{\eta}
\frac{f(\eta')-f(\eta)}{\eta-\eta'} \, d\eta'.
\end{equation}
This definition ensures that the memory integral is finite while preserving the causal dependence on the previous history of the mode. The effective Schr\"{o}dinger equation now reads
\begin{equation} \label{Corrected fractional schrodinger}
    i{}^{C}D^{\alpha}_{\eta}\psi_{\mathbf{k}}^{(1)}(\eta)=\mathcal{H}_{\mathbf{k}}\psi^{(1)}_{\mathbf{k}}+\frac{\psi_{\mathbf{k}}^{(1)}}{m_P^2\psi^{(0)}_{\mathbf{k}}}\left(\frac{1}{\gamma}\mathcal{G}_{AB}\frac{\partial \psi_{\mathbf{k}}^{(0)}}{\partial q_A}\frac{\partial \gamma}{\partial q_B}-\frac{1}{2}\mathcal{G}_{AB}\frac{\partial^2\psi_{\mathbf{k}}^{(0)}}{\partial q_A\partial q_B}\right).
\end{equation}
where 
\begin{equation}\label{Fractional dimension correction oder m_p^2}
    \alpha=1-\frac{A\psi_{\mathbf{k}}^{(1)}}{m^2_P\psi_{\mathbf{k_0}}^{(1)}}\simeq 1-\frac{A}{m^2_{P}}\left(\frac{k}{k_0}\right)^{3/4}\simeq  1-\mathcal{O}(m_P^{-2}),
\end{equation}
is the effective fractional dimension memory effects introduce. Here, for simplicity, have assumed that $A\geq 0$, so that $0<\alpha\leq 1$ and the integer $n=\lceil\alpha\rceil=1$.
The above result \footnote{We recover Caputo Derivatives in eqn.~\eqref{Corrected fractional schrodinger} rather than Riemann Lioville derivatives since the latter would require an additional $\int_0^{\eta_0}G(\eta-\eta_0)\Psi(q^A,\nu_{\mathbf{k}}) d\eta$ term in the RHS of eqn.~\eqref{Memory Wheeler De Witt}.} shows that fractional time evolution arises as an effective description of fundamentally nonlocal dynamics. Moreover, eqn.~\eqref{Fractional dimension correction oder m_p^2} shows that the fractional order generated by the memory kernel acquires a nontrivial dependence on the perturbation scale. This is a natural consequence of interpreting the kernel as a scale-sensitive memory operator, since perturbations of different wavelengths carry different histories and thus need not retain memory with the same efficiency. This results in a scale dependence modification to the primordial spectrum given by $k^{3/4}$  that we shall soon obtain. 
\\
\\
In this sense, fractional quantum cosmology does not require modifying the Wheeler-DeWitt equation at the level of differential operators. Instead, it emerges from a more fundamental nonlocal structure in which the evolution of the wave function retains information about its history.
\\
\\
In terms of the perturbation Hamiltonian, eqn.~\eqref{effective memory schrodinger} reads
\begin{equation}
            i\psi_{\mathbf{k}}^{(1)}(\eta)'=\mathcal{H}_{\mathbf{k}}\psi^{(1)}_{\mathbf{k}}-\frac{\psi_{\mathbf{k}}^{(1)}}{m_P^2\psi^{(0)}_{\mathbf{k}}}\left[ \frac{(\mathcal{H}_\mathbf{k})^2}{V}\psi_{\mathbf{k}}^{(0)}+i\frac{\partial}{\partial \eta}\left(\frac{\mathcal{H}_{\mathbf{k}}}{V}\right) \psi_{\mathbf{k}}^{(0)}\right]-\frac{iA}{m_P^2}\left(\frac{k}{k_0}\right)^{3/4}\int\frac{\psi_{\mathbf{k}}^{(0)}(\eta')'-\psi_{\mathbf{k}}^{(0)}(\eta)'}{\eta-\eta'}d\eta' .
\end{equation}
or equivalently (at order $m_P^{-2}$),
\begin{equation}
    i\space{}^{C}D_{\eta}^{\alpha}\psi_{\mathbf{k}}^{(1)}(\eta)=\mathcal{H}_{\mathbf{k}}\psi^{(1)}_{\mathbf{k}}-\frac{\psi_{\mathbf{k}}^{(1)}}{m_P^2\psi^{(0)}_{\mathbf{k}}}\left[ \frac{(\mathcal{H}_\mathbf{k})^2}{V}\psi_{\mathbf{k}}^{(0)}+i\frac{\partial}{\partial \eta}\left(\frac{\mathcal{H}_{\mathbf{k}}}{V}\right) \psi_{\mathbf{k}}^{(0)}\right].
\end{equation}
Let us now study the Gaussian ansatz for the QG-corrected Sch\"{o}dinger equation with memory dependence. We
write
\begin{equation}
    \psi_{\mathbf{k}}^{(1)}(\eta,\nu_{\mathbf{k}})=N_{\mathbf{k}}^{(1)}(\eta)e^{-\frac{1}{2}\Omega_{\mathbf{k}}^{(1)}(\eta)\nu_{\mathbf+{k}}^2},
\end{equation}
collecting all the terms with either a factor of $\nu_{\mathbf{k}}^2$ or $\nu_{\mathbf{k}}^{0}$, we have
\begin{equation}
      iN_{\mathbf{k}}^{(1) '}(\eta)=\frac{1}{2}N_{\mathbf{k}}^{(0)}(\eta)\Omega_{\mathbf{k}}^{(0)}(\eta)-\frac{A}{m_P^{2}}\left(\frac{k}{k_0}\right)^{3/4}\int \frac{N_{\mathbf{k}}^{(0)}(\eta')\Omega_{\mathbf{k}}^{(0)}(\eta')-N_{\mathbf{k}}^{(0)}(\eta)\Omega_{\mathbf{k}}^{(0)}(\eta)}{2(\eta-\eta')}d\eta,
\end{equation}
\begin{equation}
    i\Omega_{\mathbf{k}}^{(1)'}(\eta)=\left(\Omega_{\mathbf{k}}^{(0)}(\eta)\right)^2-\tilde\omega_{\mathbf{k}}^2(\eta)+\frac{iA}{m_P^{2}N_{\mathbf{k_{0}}}^{(0)}(\eta)}\int \frac{N_{\mathbf{k}}^{(0)}(\eta')\Omega^{(0)'}_{\mathbf{k}}(\eta')-N_{\mathbf{k}}^{(0)}(\eta)\Omega^{(0)'}_{\mathbf{k}}(\eta)}{(\eta-\eta')}d\eta.
\end{equation}
The above equations show that both the normalization and the Gaussian width acquire history-dependent corrections, such that their evolution at time $\eta$ depends on their entire past behavior. This represents a qualitative departure from the standard case, where the dynamics is governed by local differential equations.
\\
\\
Using eqns.\eqref{N_k^0 1/4} and \eqref{Omega 0 desitter}, we find that the additional memory dependent contribution to $\Omega_{\mathbf{k}}^{(1)'}(\eta)$ reads
\begin{equation}
    \frac{Ak^{3}}{m_P^2\eta^{1/2}}\left(\frac{k}{k_0}\right)^{3/4}\int_{0}^{\eta} \frac{\eta^{3/2}-\eta'^{3/2}}{\eta-\eta'}d\eta'=\frac{A\eta k^{15/4}}{m_P^2}\left(\frac{k}{k_0}\right)^{3/4}\left(\frac{8}{3}-2\text{ln}2\right).
\end{equation}
Finally, the memory dependent Power Spectrum now reads
\begin{equation}
    \mathcal{P}_{S,T}^{(1)}(k)=\mathcal{P}_{S,T}^{(0)}(k)\frac{\mathfrak{Re}\Omega_{\mathbf{k}}^{(1)}}{\mathfrak{Re}\Omega_{\mathbf{k}}^{(0)}}=\mathcal{P}_{S,T}^{(0)}(k)\left[1+0.988\frac{H_0}{m_P^2}\left(\frac{k_0}{k}\right)^3+\frac{A}{m_p^{2}}\left(\frac{8}{3}-2\text{ln}(2)\right)\left(\frac{k}{k_0}\right)^{3/4}\right].
\end{equation}
Hence, we see that memory effects induce a modified scale dependence of the form $k^{3/4}$, providing a distinct imprint of the underlying non-Markovian structure.
\\
\\
\begin{figure}
    \centering
    \includegraphics[width=0.8\linewidth]{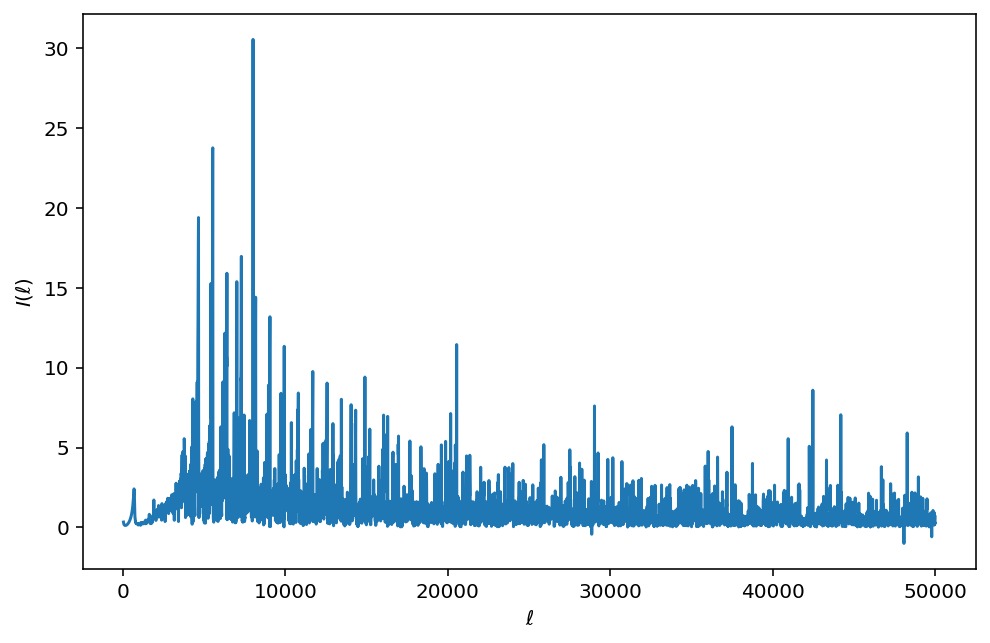}
    \caption{Numerical Plot of the integral $I(l)$ appearing in eqn.~\eqref{Temperature anisotropy at large l}, as a function of the multipole number $l$, shown for $30\leq l\leq 5\times 10^4$.}
    \label{Figure 1}
\end{figure}
This memory-sourced modification of the primordial power spectrum leads to a qualitatively new signature in the cosmic microwave background (CMB) temperature anisotropies, conventionally characterized by the angular power spectrum $\frac{l(l+1)}{2\pi}C_l$, through
\begin{equation}
    \frac{l(l+1)}{2\pi}C_l=\frac{2l(l+1)}{25}\int d\space\text{ln}k\space\Theta_l^2(k)\mathcal{P}_{S}(k)
\end{equation}
where $\Theta_l^2(k)$ denotes the transfer function.
\\
\\
In the standard Kiefer
approach, the quantum gravitational correction scales as $k^{-3}$, and therefore predominantly affects only the largest angular scales (small $l$), becoming rapidly negligible at higher multipoles. By contrast, the memory-induced correction derived in this work introduces an additional contribution proportional to $k^{3/4}$ at horizon crossing, corresponding to a blue-tilted enhancement of power toward smaller angular scales
\\
\\
For low multipoles, this correction produces a modified Sachs-Wolfe contribution
\begin{equation}\label{Temperature anisotropy corection at small l}
    \Delta C^{\text{SW}}_l\simeq  \frac{4\pi^{2} A  _s A(k_0\eta_{\text{rec}})^{-3/4}2^{-9/4}\Gamma(5/4)\Gamma(l+3/8)}{25m_P^2\Gamma^2(25/8)\Gamma(l+13/8)}\left(\frac{8}{3}-2\text{ln}(2)\right)\left(\frac{k}{k_0}\right)^{3/4}
\end{equation}
From eqn.~\eqref{Temperature anisotropy corection at small l}, we infer that, for small $l$, the memory-induced correction is generally subdominant, and the observed anisotropies remain overwhelmingly controlled by the standard inflationary power spectrum. In this regime, the phenomenology associated with memory effects is comparatively modest and difficult to disentangle from the usual large-scale contributions.
\\
\\
The most distinctive and phenomenologically significant consequences of memory effects arise instead at high multipoles. 
\begin{equation}\label{Temperature anisotropy at large l}
\begin{split}
    \Delta\frac{l(l+1)}{2\pi}C_l=\frac{A_s A    }{m_P^2(k_{0}\eta_{\text{rec}})^{3/4}}\frac{16-12\text{ln}2}{15}\\
    \int_1^\infty \frac{d\beta}{\beta^2\sqrt{\beta^2-1}}\left[F^2\left(\frac{l\beta}{|\eta_{rec}-\eta_{horizon}|}\right)+\frac{\beta^2-1}{\beta^2}G^2\left(\frac{l\beta}{|\eta_{rec}-\eta_{horizon}|}\right)\right](l\beta)^{3/4}\\=\frac{A_s A}{m_P^2(k_{0}\eta_{\text{rec}})^{3/4}}\frac{8-6\text{ln}2}{3} I_{l}
\end{split}
\end{equation}
From eqn.~\eqref{Temperature anisotropy at large l} we see this at $l\gtrsim 30$. These effects are more pronounced especially in the range of several thousand as the blue-tilted $k^{3/4}$ correction generates an enhancement of power toward small angular scales, culminating in pronounced oscillatory peaks as shown in Fig.~(\ref{Figure 1}), with the strongest feature occurring near $l\approx 6250$. This behavior is qualitatively different from the standard semiclassical quantum gravitational correction, which is important primarily at low $l$ and becomes strongly suppressed at smaller angular scales. The emergence of a characteristic high $l$ excess therefore constitutes the principal observational signature of the memory-dependent effects developed in this work and provides a uniquely identifiable imprint of non-Markovian quantum gravitational dynamics.
\\
\\ 
If the effective memory strength satisfies
\begin{equation} \label{ampbound}
\frac{A}{m_P^2}\gtrsim 10^{-3},
\end{equation}
the resulting contribution can become comparable to the standard temperature anisotropies at the corresponding high multipoles. This estimate should not be interpreted as a sharp observational bound, but rather as an order of magnitude threshold indicating when the memory-induced term becomes phenomenologically relevant. The parameter $A/m_P^2$ is dimensionless in the present normalization and controls the deviation of the effective fractional order from unity, $\alpha=1-A/m_P^2$ and so for $A/m_P^2\sim 10^{-3}$, the departure from ordinary first order time evolution remains small, with $\alpha$ still extremely close to one. In this sense, the correction is still compatible with the perturbative semiclassical treatment provided that the memory contribution to the mode functions remains subdominant to the leading Schrödinger evolution over the range of scales under consideration. The naturalness of such a value depends on the microscopic origin of the memory kernel. We say this because, although one might naively expect $A/m_P^2$ to be very small when $A$ is generated by integrating out Planck-scale degrees of freedom, nonlocal memory effects can accumulate over long conformal time intervals. As a result, the relevant parameter in observables may not simply be the local Planck suppression. It could instead be the integrated strength of the kernel along the post-nucleation history, hence, values of order of magnitude $10^{-3}$ need not automatically be regarded as inconsistent, although they should be viewed as phenomenologically significant rather than generic. 
\\
\\
Importantly, this regime does not by itself signal a breakdown of the $m_P^{-2}$ expansion as the expansion would become unreliable only if the memory-induced correction to the power spectrum became comparable to, or larger than, the leading order spectrum over a broad range of modes. It could also become unreliable if higher order memory terms of order $m_P^{-4}$ and beyond were no longer suppressed. Hence the perturbative description remains self consistent as long as the correction is treated as a small deformation of the standard spectrum and the hierarchy of
\begin{equation}
\left|\Delta P_{\rm mem}(k)\right| \ll P^{(0)}(k)
\end{equation}
is maintained over the range where the calculation is applied. At the same time, such a value of $A/m_P^2$ would be strongly constrained as well. It would be so because the memory term is blue tilted and enhances short scale power, it can be tested not only by high-$l$ CMB temperature anisotropies but also by CMB damping-tail measurements\cite{Keisler_2011,Story_2013}, CMB lensing \cite{2020,Qu_2024,Wu_2019,Qu_2026} etc. This would mean that the estimate above should be regarded as an observational target rather than a freely allowed parameter range. A full constraint on $A$ would require evolving the modified primordial spectrum through a Boltzmann code such as CLASS \cite{lesgourgues2011cosmiclinearanisotropysolving} or CAMB \cite{Lewis_2000}, including foregrounds, Silk damping, lensing and nonlinear small scale effects \cite{2020,2020v,2020vii}. Such an analysis lies beyond the scope of the present work, but it provides a direct route for converting the memory coefficient into a measurable or constrainable parameter.
\\
\\
Furthermore, if the memory parameter $A$ is dynamically selected or finely tuned to values near the observational threshold, the resulting enhancement of power at specific small scales could have important consequences for structure formation. In particular, the localized amplification of primordial fluctuations may influence the abundance and distribution of bound structures across a wide range of scales, potentially affecting the formation of galaxies, planetary systems, and, indirectly, environments capable of supporting intelligent life.
\section{Implications of Primordial Non-Gaussianity } \label{Non-Gaussianity}
A natural question to ask is, given the effects of the memory paradigm on power spectra, could one envisage effects on primordial non-Gaussianity signatures too \cite{png11takahashi2014primordial,png1meerburg2019primordial,png2pajer2012new}? This question is natural because the same memory kernel that modifies the two point function should, also possibly affect higher order correlation functions. The memory-induced correction to the primordial power spectrum is scale dependent and so it leads to a departure from the standard power spectrum. Since primordial non-Gaussianity is sensitive to the time evolution of modes, their interactions and the correlations between different wavelengths, we can expect that the nonlocal memory structure could leave imprints not only in $P_\zeta(k)$, but also in the bispectrum $B_\zeta(k_1,k_2,k_3)$. We now investigate this possibility in more detail. 
\\
\\
Let us write the memory corrected scalar power spectrum in the schematic form \cite{png3munshi2010new,png4scoccimarro2004probing,png5floss2023primordial,png6verde2001tests}
\begin{equation}
P_\zeta(k,\eta) = P_\zeta^{(0)}(k) \left[
1+\delta_{\rm mem}(k,\eta) \right],
\end{equation}
note that $P_\zeta^{(0)}(k)$ denotes the standard leading order spectrum and $\delta_{\rm mem}(k,\eta)$ denotes the correction induced by memory effects. From the result obtained in the previous section, the memory contribution has the form
\begin{equation}
\delta_{\rm mem}(k,\eta)= C_{\rm mem}
\frac{A}{m_P^2}  \left(\frac{k}{k_0}\right)^{3/4}
\end{equation}
where
\begin{equation}
C_{\rm mem}= 
\frac{8}{3} - 2\ln 2  \simeq 1.28
\end{equation}
This expression is important for us because it already shows the qualitative nature of the expected non-Gaussian signal. It is so because it shows that the memory correction is blue tilted and becomes more relevant at large $k$. Therefore, any associated non-Gaussianity should also be expected to be strongest at small scales rather than at the largest CMB angular scales. A first estimate can be obtained from the squeezed limit, where one mode is much longer than the other two, $k_L\ll k_S$ and in that case we can write squeezed bispectrum as
\begin{equation}
B_\zeta(k_L,k_S,k_S) \simeq \frac{12}{5}
f_{\rm NL}^{\rm sq} P_\zeta(k_L)P_\zeta(k_S)
\end{equation}
In standard single field inflation the squeezed-limit consistency relation is the famous Maldacena one \cite{maldacena2003non}
\begin{equation}
f_{\rm NL}^{\rm sq}= \frac{5}{12}(1-n_s)
\end{equation}
In the present case, the effective scalar spectral index receives an additional memory-induced contribution and as 
\begin{equation}
n_s^{\rm eff}(k)-1=
\frac{d\ln P_\zeta(k)}{d\ln k}
\end{equation}
we obtain
\begin{equation}
n_s^{\rm eff}(k)-1= n_s^{(0)}-1 +
\frac{d\ln\left(1+\delta_{\rm mem}\right)}{d\ln k}
\end{equation}
For $\delta_{\rm mem}\ll 1$, this becomes
\begin{equation}
\frac{d\ln\left(1+\delta_{\rm mem}\right)}{d\ln k}
\simeq \frac{d\delta_{\rm mem}}{d\ln k}= \frac{3}{4}\delta_{\rm mem}
\end{equation}
So, if the squeezed limit consistency logic remains approximately valid then we can write the memory-induced correction to the squeezed non-Gaussianity to be given as \cite{png7ferrante2022primordial,png8seljak2009extracting}
\begin{equation}
\Delta f_{\rm NL}^{\rm sq}
\simeq -\frac{5}{12}
\frac{d\delta_{\rm mem}}{d\ln k}= -\frac{5}{16}\delta_{\rm mem}
\end{equation}
Using the explicit form of $\delta_{\rm mem}$, this gives us
\begin{equation}
\Delta f_{\rm NL}^{\rm sq} \simeq -\frac{5}{16}
C_{\rm mem} \frac{A}{m_P^2}  \left(\frac{k_S}{k_0}\right)^{3/4}
\end{equation}
Numerically, this can be written as
\begin{equation}
\Delta f_{\rm NL}^{\rm sq} \simeq -0.4 \frac{A}{m_P^2}  \left(\frac{k_S}{k_0}\right)^{3/4}
\end{equation}
This estimate is conservative, since it assumes that memory affects the squeezed bispectrum only through the modified scale dependence of the power spectrum. Even in this minimal treatment, however, the main point is clear: the non-Gaussian correction inherits the same blue scaling as the memory correction to the power spectrum. The squeezed signal is therefore not expected to be largest near the usual CMB pivot scales, but rather at shorter wavelengths, where the memory contribution becomes more important.
\\
\\
There is also a conceptual issue with applying the usual squeezed-limit consistency relation too directly. In the standard argument, the long-wavelength mode acts mainly as a local rescaling of the short-wavelength background. In the presence of memory, this assumption can be weakened, because the short modes retain information about their past evolution. The long mode may then modulate not only the instantaneous local background, but also the accumulated history that enters the memory integral. A more general squeezed-limit parametrization can therefore be written as \cite{png11takahashi2014primordial,png13giannantonio2012constraining,png14bartolo2005signatures}
\begin{equation}
f_{\rm NL}^{\rm sq,mem}(k_L,k_S) = f_{\rm NL}^{\rm sq,0} + \mathcal{C}_{\rm sq} \frac{A}{m_P^2}
\left(\frac{k_S}{k_0}\right)^{3/4}
\mathcal{F}_{\rm sq} \left(\frac{k_L}{k_S}\right),
\end{equation}
where $\mathcal{C}_{\rm sq}$ is a model-dependent coefficient, while $\mathcal{F}_{\rm sq}(k_L/k_S)$ describes the shape dependence induced by the memory kernel. If the memory term produces genuine long-short temporal correlations, then $\mathcal{F}_{\rm sq}$ need not reduce to the standard single field form. A departure from the usual squeezed-limit behavior would therefore be a useful indication that the perturbations are not governed by purely local in time evolution. One may also estimate the effect in equilateral configurations, where $k_1=k_2=k_3=k$, for which we note that the equilateral bispectrum is written as
\begin{equation}
B_\zeta(k,k,k)= \frac{18}{5} f_{\rm NL}^{\rm eq}
P_\zeta^2(k)
\end{equation}
If the only role of memory were to correct the external mode functions, then the mode functions would transform approximately as
\begin{equation}
u_k \rightarrow u_k \left( 1+\frac{1}{2}\delta_{\rm mem} \right).
\end{equation}
Since the bispectrum contains three external legs, this would imply
\begin{equation}
B_\zeta \rightarrow B_\zeta^{(0)} \left(
1+\frac{3}{2}\delta_{\rm mem} \right).
\end{equation}
On the other hand, the denominator entering $f_{\rm NL}^{\rm eq}$ contains two powers of the power spectrum, so that
\begin{equation}
P_\zeta^2 \rightarrow \left(P_\zeta^{(0)}\right)^2
\left( 1+2\delta_{\rm mem} \right).
\end{equation}
So, the purely external leg correction gives
\begin{equation}
f_{\rm NL}^{\rm eq} \rightarrow f_{\rm NL}^{\rm eq,0}
\left( 1-\frac{1}{2}\delta_{\rm mem} \right)
\end{equation}
or equivalently we can also have
\begin{equation}
\Delta f_{\rm NL}^{\rm eq} \sim -\frac{1}{2}
f_{\rm NL}^{\rm eq,0} \delta_{\rm mem}(k)
\end{equation}
This contribution is small if the original slow roll value of $f_{\rm NL}^{\rm eq,0}$ is small. So, what we realize here is that the most important equilateral signal is not expected to arise merely from external-leg renormalization. It would instead arise from direct memory induced cubic terms, because the same nonlocal structure that modifies the quadratic dynamics should also appear in the cubic evolution of perturbations. A phenomenological cubic memory contribution can be taken as 
\begin{equation}
S_3^{\rm mem}
\sim \frac{A}{m_P^2} \int d\eta,d\eta'\,
a^2(\eta) K(\eta-\eta') \zeta^2(\eta)
\partial_{\eta'}\zeta(\eta').
\end{equation}
This expression is not meant to be the unique cubic action, but rather the simplest schematic term showing how a nonlocal kernel can correlate perturbations evaluated at different times. The associated in-in contribution to the bispectrum takes the general form
\begin{equation}
B_\zeta^{\rm mem} = -2,{\rm Im}
\left[ u_{k_1}(0)u_{k_2}(0)u_{k_3}(0)
\int d\eta,d\eta' \,
K(\eta-\eta') \mathcal{I}_{k_1k_2k_3}(\eta,\eta')
\right]\end{equation}
where $\mathcal{I}_{k_1k_2k_3}(\eta,\eta')$ contains the products of mode functions and their derivatives appropriate to the interaction under consideration. Note again that we used a kernel with a plus-distribution structure given in eqn.\eqref{kernel}, the corresponding time integrals naturally generate logarithmic dependence on the triangle perimeter $K_t=k_1+k_2+k_3$ and so we have
\begin{equation}
\int_0^\eta \left[ \frac{1}{\eta-\eta'}
\right]_+ e^{iK_t\eta'} d\eta' \sim -\ln(K_t|\eta|)
-\gamma_E -\frac{i\pi}{2} +\cdots 
\end{equation}
Thus, a rough equilateral memory template is expected to take the form
\begin{equation}
f_{\rm NL}^{\rm eq,mem}(k) \sim
\mathcal{C}_{\rm eq} \frac{A}{m_P^2}
 \left(\frac{k}{k_0}\right)^{3/4}
\ln\left(\frac{k}{k_{\rm nuc}}\right),
\end{equation}
where $\mathcal{C}_{\rm eq}$ is a model dependent coefficient and $k_{\rm nuc}$ is an effective scale associated with the lower limit of the memory integral, or equivalently with the nucleation surface. This result shows that the equilateral signal may contain both a blue running and a logarithmic enhancement and so such a structure would be qualitatively different from the nearly scale independent non-Gaussianity expected in minimal slow roll scenarios.
\\
\\
The memory kernel can also generate signatures in folded configurations and this is because nonlocal in-time interactions are sensitive to phases accumulated over the history of the mode functions. In ordinary Bunch-Davies calculations, the dominant oscillatory phases usually involve combinations of the form
\begin{equation}
e^{-i(k_1+k_2+k_3)\eta}
\end{equation}
However, when memory terms correlate earlier and later times or effectively generate a small admixture of negative-frequency evolution, one may obtain phase combinations such as $k_1+k_2-k_3,
k_1-k_2+k_3 $ and $
-k_1+k_2+k_3$ and these combinations become enhanced in flattened limits, for example when
\begin{equation}
k_1+k_2\simeq k_3
\end{equation}
So one expects a folded memory contribution of the schematic form
\begin{equation}
B_\zeta^{\rm fold,mem}
\propto \frac{A}{m_P^2}
\left(\frac{K}{k_0}\right)^{3/4}
P_\zeta(k_1)P_\zeta(k_2)
\ln\left| \frac{k_1+k_2+k_3}
{k_1+k_2-k_3} \right| +{\rm perms.}
\end{equation}
This folded enhancement could be one of the cleanest non-Gaussian signatures of the memory paradigm. Standard single field slow roll inflation does not usually produce a sharply enhanced folded bispectrum, but nonlocal memory effects naturally contain the phase structure needed to amplify flattened triangle configurations. The scale dependence of the resulting non-Gaussianity can be characterized through the running parameter
\begin{equation}
n_{\rm NG}^{\rm mem} = \frac{d\ln |f_{\rm NL}^{\rm mem}|}{d\ln k}
\end{equation}
If the dominant memory contribution scales as
\begin{equation}
f_{\rm NL}^{\rm mem}(k) \propto k^{3/4}
\ln(k/k_{\rm nuc})
\end{equation}
then its running is
\begin{equation}
n_{\rm NG}^{\rm mem}= \frac{3}{4} + \frac{1}{\ln(k/k_{\rm nuc})}
\end{equation}
So up to logarithmic corrections, one obtains the approximate prediction
\begin{equation}
n_{\rm NG}^{\rm mem} \simeq 0.75
\end{equation}
This is a rather strong scale dependence compared with the nearly scale-independent non-Gaussianity expected in minimal slow roll inflation. It also means that even if the memory induced bispectrum is small at the usual CMB pivot scale, it may grow significantly at smaller scales where the memory correction to the power spectrum is also enhanced.
\\
\\
Finally, to obtain a rough estimate of the amplitude, we can suppose that the memory coefficient lies near the phenomenologically relevant threshold given by eqn.~\eqref{ampbound} and so, the memory correction becomes observationally relevant only after the scale dependent enhancement factor $\left(\frac{k}{k_0}\right)^{3/4}$ is sufficiently large. If at a given high $k$ scale, the net fractional power spectrum correction is \begin{equation}
\delta_{\rm mem}(k) \sim 10^{-2}
\end{equation}
then the conservative squeezed estimate gives
\begin{equation}
|\Delta f_{\rm NL}^{\rm sq}| \sim \frac{5}{16}\delta_{\rm mem} \sim 3\times 10^{-3}
\end{equation}
Such a signal would likely be too small to be directly observed in the ordinary large scale CMB bispectrum \cite{planckpngade2014planck}. However, this estimate only captures the indirect contribution associated with the modified tilt of the power spectrum and it in no way includes direct cubic memory interactions, which can be parametrically more important because they probe the nonlocal kernel itself. If direct cubic memory terms contribute with an order one coefficient, one may instead estimate
\begin{equation}
f_{\rm NL}^{\rm mem} \sim \mathcal{C} \delta_{\rm mem}(k) \ln(k/k_{\rm nuc})
\end{equation}
For
\begin{equation}
\delta_{\rm mem} \sim 10^{-2}-10^{-1},
\qquad \ln(k/k_{\rm nuc})
\sim 10-50
\end{equation}
this gives the rough range
\begin{equation}
f_{\rm NL}^{\rm mem} \sim 10^{-1}-{\rm few}.
\end{equation}
This is a slightly more phenomenologically interesting regime because it suggests that the ordinary large scale CMB bispectrum may remain consistent with current bounds, while small scale observables such as $\mu$-distortion anisotropy correlations \cite{Pajer_2012,Chluba:2012gq}, primordial black hole statistics \cite{Pi_2025,ATAL2019100275} and small scale structure could carry significantly larger memory induced non-Gaussian signatures.
\section{Fine-tuning of the coefficient $A$}
\label{self}
A particularly important feature of the memory-induced corrections derived in Section (\ref{frac KK-f}) is that they predominantly affect large multipoles $l$, corresponding to smaller angular scales, unlike the standard semiclassical quantum gravitational corrections, which mainly modify the small $l$ sector and rapidly become negligible large multipoles.
\\
\\
These small angular scale fluctuations are important because they feed directly into the later formation of cosmological structure. The high $l$ part of the primordial spectrum controls the small scale inhomogeneities that eventually seed galaxies, stellar systems, compact objects, and planetary systems. The memory coefficient $A$ therefore affects not just a formal correction to the power spectrum, but also the amplitude and distribution of the density fluctuations that enter structure formation. If $A$ is very small, the memory contribution is essentially negligible and the theory reduces, for practical purposes, to the standard semiclassical picture. If $A$ is too large, however, the enhancement of short scale power could become strong enough to change structure formation in an undesirable way, for example by producing excessive small scale clustering or a strongly non standard astrophysical history. This suggests that only some range of values of $A$ may lead to long-lived gravitationally bound structures, stable stellar environments, complex chemistry, and eventually observers. In this sense, $A$ acts as a cosmological parameter controlling how efficiently small scale structure forms and how stable that structure can be.
\\
\\
This immediately raises a conceptual issue. In the framework developed here, memory effects become physically meaningful only after the universe enters the classically allowed Lorentzian regime at the nucleation surface $a=H^{-1}$. Before this transition, the cosmological wave function is described by a four-dimensional Euclidean geometry, not by a causal spacetime history in which memory can accumulate. Thus, within a single cosmological history, there is no obvious mechanism by which the parameter $A$ could be gradually driven toward phenomenologically preferred values. The issue is therefore not only parameter selection, but also the origin of cosmological memory itself: if the universe has only a finite post-nucleation history, then the appearance of a finely tuned memory coefficient remains unexplained.
\\
\\
One possible way around this is to place the present framework inside a cyclic cosmological setting \cite{doi:10.1126/science.1070462,LEHNERS2008223,PhysRevD.86.083536}, where the universe can inherit information from earlier cosmological cycles. In such a picture, the effective memory coefficient would not have to be a fixed fundamental number. It could instead change from one cycle to the next, through the cumulative effect of memory carried across successive cosmological histories. A natural setting for this idea is conformal cyclic cosmology (CCC), originally proposed by Roger Penrose \cite{Markwell2022TowardFA,penrose2010cycles,Penrose:2006zz,Penrose:2014vok,meissner2025physicsconformalcycliccosmology}. In CCC the observed universe is only one cycle, or aeon, in an infinite sequence of cosmological epochs, rather than a single isolated spacetime history emerging from one final Big Bang. The exponentially expanded late-time future of one aeon is conformally identified with the Big Bang geometry of the next \cite{Penrose:1964ge}, so that spacetime continues through an infinite chain of conformally connected phases.
\\
\\
In the standard Hawking-Hartle no-boundary proposal, by contrast, the universe is described by a Euclidean gravitational path integral over compact geometries,
\begin{equation}
\Psi[h_{ij},\phi]=\int_{g|_{\partial\mathcal{M}}=h^2, \Phi|_{\partial\mathcal{M}}=\phi}\mathcal{D}g\mathcal{D}\phi e^{-S_E[g,\phi]/\hbar}.
\end{equation}
where $S_{E}$ is the Euclidean action and the integration is taken over regular compact four-geometries inducing the boundary data $(h_{ij},\phi)$. In this framework, classical Lorentzian spacetime emerges only after the nucleation surface $a=H^{-1}$, where the Euclidean geometry smoothly transitions into an expanding Lorentzian universe.
\\
\\
Motivated by the complementary features of these two frameworks, we consider a cyclic extension of the Hawking-Hartle proposal \footnote{Here, we focus only on a cyclic extension of the Hawking-Hartle proposal, as its implementation is conceptually simpler and involves fewer subtleties than a cyclic extension of the Vilenkin-Yamada proposal, which we leave for future work. A potential concern is that the Vilenkin-Yamada wave function generically favors inflationary initial conditions, whereas inflation and cyclic cosmology have traditionally been regarded as competing paradigms. However, recent essay \cite{jay2026cosmologicalinflationinsidecyclic} has argued that inflationary dynamics may arise naturally within a cyclic cosmological framework, suggesting that the two scenarios need not be mutually exclusive.} in which the conformal crossover between successive aeons is identified with the boundary $\partial{\mathcal{M}}$ of the Euclidean instanton itself. In this picture, the Euclidean sector no longer describes an isolated nucleation event for a single universe, but instead acts as the conformal bridge connecting the asymptotic future of one aeon to the Lorentzian emergence of the next. The Hawking-Hartle geometry therefore acquires an intrinsically cyclic interpretation, with the nucleation hypersurface simultaneously serving as an aeonic crossover surface.
\\
\\
In the cyclic extension proposed here, the Hawking-Hartle geometry acquires a different global interpretation. Each aeon is still assumed to originate from a Hawking-Hartle-type nucleation process, with Lorentzian spacetime emerging when the scale factor reaches $a=H^{-1}$. However, unlike the standard no-boundary proposal, this nucleation event is not interpreted as the absolute beginning of spacetime. Instead, the asymptotic conformal future of a preceding Lorentzian aeon is identified directly with the nucleation hypersurface of the subsequent aeon. The conformal crossover therefore occurs between successive Lorentzian cosmological phases, while the associated Euclidean Hawking-Hartle geometry specifies the nucleation conditions for each aeon.
\\
\\
Schematically, the global structure takes the form
\begin{equation}
    \mathcal{A}_n^{\text{Lor}}\rightarrow \mathcal{A}_{n+1}^{\text{Lor}}
\end{equation}
with each aeon individually satisfying a Hawking-Hartle type nucleation condition at $a=H^{-1}$. In this picture, the boundary $\partial \mathcal{M}$ of the Euclidean instanton no longer represents an isolated origin of spacetime, but instead defines the crossover hypersurface separating consecutive aeons. The universe therefore possesses no unique initial cosmological epoch; rather, the Hawking-Hartle nucleation geometry recurs cyclically through an infinite sequence of conformally connected cosmological phases. 
\\
\\
Within this cyclic Hawking-Hartle framework, the coefficient $A$ appearing in the memory kernel is no longer interpreted as a fixed fundamental constant. Instead, it becomes a parameter that may evolve from aeon to aeon through the cumulative inheritance of correlations across successive cosmological cycles. Physically, $A$ characterises the strength with which the universe retains information about its prior dynamical history after the onset of the Lorentzian regime at $a=H^{-1}$. Since the memory-induced corrections derived in Section (\ref{frac KK-f}) predominantly affect the high-$l$ sector of the primordial spectrum, the value of $A$ directly influences the formation of small scale cosmological structure, including the distribution of galaxies, stellar systems, and other gravitationally bound configurations. Consequently, different aeons may exhibit qualitatively different structure-formation histories depending on the effective value of the inherited memory coefficient.
\\
\\
This suggests that cyclic cosmology may provide a mechanism through which the parameter $A$ undergoes a  progressive dynamical refinement across successive aeons. To make this idea more concrete, suppose that the existence of an aeon capable of supporting complex structures and intelligent observers requires the memory coefficient to approach a distinguished value $A_0$ determined implicitly through a constraint of the form
\begin{equation}
    G(A_0,\beta_1,\beta_2,\dots)=0
\end{equation}
where $\beta_i$ are other fundamental parameters of the theory. Although the precise form of $G$ is highly complex and non-trivial, it schematically represents physical conditions necessary for the formation of stable structures and cosmological complexity.
n this picture, the sequence of aeons induces an effective evolution 
\begin{equation}
    A_n\rightarrow A_0 \quad \text{as $n\rightarrow \infty$}
\end{equation}
where $A_n$ denotes the memory coefficient associated with the $n$-th aeon and the parameter $A_0$ is the memory coefficient required to support complex structures and intelligent observers.

\section{Conclusions}
\label{conc}
In this work, we have developed a way of incorporating memory effects into quantum cosmology by extending the usual semiclassical expansion of the Wheeler-DeWitt equation beyond its standard local form. Starting from the 
approach in \cite{PhysRevD.93.104035,Brizuela:2016gnz,Brizuela:2019jzv,PhysRevD.44.1067,Barvinsky_1998,PhysRevD.72.045006,Kiefer:2011cc}, we first reviewed how the classical background dynamics, the emergent Schr\"{o}dinger equation for perturbations and the first quantum gravitational corrections arise in an expansion in powers of the Planck mass. This gives the standard reference point for the rest of the paper, since any nonlocal modification has to preserve this semiclassical hierarchy at leading order. We then considered whether nonlocality could be introduced more directly by replacing ordinary derivatives with fractional derivatives \cite{Oldham1974TheFC,Samko1993FractionalIA,math13223643,liouville1832memoire,Riemann1876Versuch,podlubny1998fractional,kilbas2006theory,10.1111/j.1365-246X.1967.tb02303.x,caputo1969elasticita,Riesz1949}. The main lesson is that this simple fractionalization does not work in a clean way and in particular, replacing the derivatives in the Wheeler-DeWitt equation or in the Kiefer equations by Caputo derivatives \cite{10.1111/j.1365-246X.1967.tb02303.x,caputo1969elasticita}, which are usually more convenient than Riemann-Liouville derivatives for initial-value problems, is not compatible with the WKB expansion. The reason is that fractional derivatives are intrinsically nonlocal and do not act on the WKB ansatz in the same simple way as ordinary derivatives. As a result, the expansion in powers of $m_P$ no longer separates cleanly, and one does not obtain a closed Schr\"{o}dinger equation for the perturbation modes.
\\
\\
For this reason, we introduced memory effects at the level of the Wheeler-DeWitt equation itself, through a causal memory kernel that produces nonlocal, history-dependent corrections and we treated this term as a subleading contribution of order $m_P^{-2}$, so that the leading Hamilton-Jacobi and Schr\"{o}dinger equations remain unchanged. In this formulation, the dynamics is genuinely non-Markovian but still organized within the usual semiclassical expansion. Fractional time evolution then appears not as something imposed by hand, but as an effective description of the underlying memory kernel, with the fractional order measuring the strength of the history dependence.
\\
\\
We applied this framework to cosmological perturbations in a de Sitter background and derived the corresponding modified Schr\"{o}dinger equation and using a Gaussian ansatz, we found that the normalization and Gaussian width no longer obey purely local differential equations. Instead, they satisfy integro-differential equations whose values at a given time depend on the previous history of the mode. Even with this nonlocality, the corrections can still be treated perturbatively in powers of $m_P^{-1}$, and their physical meaning remains transparent. The main result is the modification of the primordial power spectrum as alongside the standard quantum gravitational correction proportional to $k^{-3}$, the memory kernel considered here generates a new contribution with a characteristic $k^{3/4}$ scale dependence. This provides a distinct imprint of the non-Markovian structure. If such a term is present, it would lead to departures from the usual quantum cosmological corrections to inflationary predictions, especially on smaller scales.
\\
\\
It should be kept in mind, however, that this particular scaling is not universal as different memory kernels can give different corrections to the power spectrum, with their own scale and time dependence. In some cases, the order $m_P^{-2}$ correction may not even freeze after horizon crossing, since the memory term can leave an explicit $\eta$ dependence on super-horizon scales. Such kernels may also fail to admit a simple effective fractional description, at least not one of the Caputo type. Another point concerns the memory coefficient $A$ discussed in Section (\ref{self}). Since the memory induced correction contributes as $k^{3/4}$, its largest observational effects appear at high multipoles $l$ or equivalently at small angular scales. The correction is hence expected to be relatively weak at low $l$, where the CMB anisotropies are dominated by the standard inflationary spectrum, but it can become more relevant in the high-$l$ regime. These short-wavelength perturbations are also important for structure formation, since they seed galaxies, stellar systems, compact objects and planetary environments. In that sense, the value of $A$ can influence the efficiency and character of structure formation and this suggests to us that only a limited range of values may lead to stable astrophysical environments, and possibly to conditions suitable for complex life. The apparent tuning of $A$ then raises the question of whether there could be a dynamical selection mechanism. One possible setting is the cyclic Hawking-Hartle picture discussed here, where memory-dependent correlations can be inherited across successive cosmological aeons. In such a scenario, $A$ need not be a fixed parameter chosen once and for all, but may evolve through cumulative inter-aeonic memory effects toward values compatible with the observed universe.
\\
\\
The same enhancement of small scale power can also have implications for PBHs \cite{Zeldovich:1967lct,10.1093/mnras/168.2.399,Carr:1975qj,Sasaki:2025frv}, as PBH formation is exponentially sensitive to the amplitude of primordial fluctuations on small scales, the blue-tilted $k^{3/4}$ correction found here could increase the PBH abundance in certain mass windows. This gives another possible observational handle on the memory based quantum gravitational dynamics developed in this work. A detailed study of memory-induced PBH production, together with the corresponding observational constraints, will be left for future work.
\\
\\
We also explored the implications of the memory paradigm for primordial non-Gaussianity \cite{png11takahashi2014primordial,png1meerburg2019primordial,png2pajer2012new}. Since the same nonlocal kernel that modifies the two-point correlation function can also affect higher order correlations, memory effects are expected to generate distinctive signatures in the primordial bispectrum. At the phenomenological level, we found that the induced non-Gaussianity inherits the same blue-tilted scaling that characterizes the memory correction to the power spectrum, leading to enhanced signals on small scales rather than at the conventional CMB pivot scales. In the squeezed limit, this produces scale-dependent corrections to $f_{NL}$, while more general memory-induced long-short temporal correlations may lead to departures from the standard Maldacena consistency relation \cite{maldacena2003non}. We further argued that nonlocal-in-time interactions naturally generate equilateral and folded bispectrum contributions with characteristic  $k^{3/4}$ running and logarithmic scale dependence. In particular, a pronounced folded enhancement would constitute a potentially clean observational signature of memory-driven non-Markovian dynamics, distinguishing this framework from conventional single field slow roll inflation. Our estimates suggest that primordial non-Gaussianity provides an additional and potentially powerful observational probe of memory-induced quantum gravitational effects.
\\
\\
More broadly, our results establish a conceptual link between semiclassical quantum cosmology, nonlocal dynamics, and fractional calculus. Rather than postulating fractional evolution at the outset, we have shown that it can arise as an effective description of a more fundamental memory-dependent structure. This clarifies both the origin and the domain of validity of fractional approaches in quantum gravity.

\section*{Acknowledgments} 
P.M. acknowledges the FCT grant UID/212/2025 CMA-UBI plus the COST Actions CA23130 (Bridging high and low energies in search of quantum gravity
(BridgeQG)) and CA23115 (Relativistic Quantum Information (RQI)). The
work of O.T. was supported in part by the Vanderbilt Discovery Doctoral
Fellowship.
\bibliographystyle{unsrt}
\bibliography{references}
\end{document}